\documentclass{aa}
\usepackage{psfig}
\usepackage{natbib}
\usepackage{longtable}
\usepackage{graphicx}
\bibpunct{(}{)}{,}{a}{}{,} 

\def\la{\;
\raise0.3ex\hbox{$<$\kern-0.75em\raise-1.1ex\hbox{$\sim$}}\; }
\def\ga{\;
\raise0.3ex\hbox{$>$\kern-0.75em\raise-1.1ex\hbox{$\sim$}}\; }

\newcommand{\kms}{km~s$^{-1}\,$}
\newcommand{\ms}{m~s$^{-1}\,$}
\newcommand{\cm}{cm$^{-2}\,$}

\newcommand{\etal}{{et al.}}
\newcommand{\zabs}{$z_{\rm abs}$}
\newcommand{\mga}{$^{24}$Mg}
\newcommand{\mgb}{$^{25}$Mg}
\newcommand{\mgc}{$^{26}$Mg}

\begin{document}

\title{First measurement of Mg isotope abundances at high redshifts and 
accurate estimate
of $\Delta\alpha/\alpha$\thanks{Based on
observations performed at the VLT Kueyen telescope (ESO, Paranal, Chile),
the ESO programme No. 183.A-0733}
}
\author{I. I. Agafonova\inst{1,2}
\and
P. Molaro\inst{1}
\and
S. A. Levshakov\inst{2,3}
\and
J. L. Hou\inst{3}
}
\institute{
INAF-Osservatorio Astronomico di Trieste, Via G. B. Tiepolo 11,
34131 Trieste, Italy
\and
Ioffe Physical-Technical Institute, 
Polytekhnicheskaya Str. 26, 194021 Saint Petersburg, Russia
\and
Key Laboratory for Research in Galaxies and Cosmology,
Shanghai Astronomical Observatory, CAS, 80 Nandan Road, Shanghai 200030,
P.R. China
}
\date{Received 00  ; Accepted 00}
\abstract
{}
{
Abundances of Mg isotopes $^{24}$Mg, $^{25}$Mg, and $^{26}$Mg can be used to test 
models of chemical enrichment of interstellar/intergalactic gas clouds. 
Additionally, since the position of the \ion{Mg}{ii} $\lambda\lambda2796, 2803$ \AA\ lines
is often taken as
a reference in computations of possible changes of the fine-structure
constant $\alpha$, it should be clarified to what extent these lines are
affected by isotopic shifts. 
} 
{
We use a high-resolution spectrum (pixel size $\approx$1.3 \kms) of the quasar
\object{HE0001--2340} observed with the UVES/VLT to measure Mg isotope abundances in the
intervening absorption-line systems at high redshifts. 
Line profiles are prepared accounting for possible shifts between the individual exposures.
In the line fitting procedure, the lines of each ion are treated independently.
Due to unique composition of the selected systems~-- the presence of several transitions
of the same ion~-- we can test the local accuracy of the wavelength scale calibration
which is the main source of errors in the sub-pixel line position measurements.
} 
{ 
In the system at \zabs\ = 0.45 which is probably
a fragment of the outflow caused by SN Ia 
explosion of high-metallicity white
dwarf(s) we measured velocity shifts of \ion{Mg}{ii} and \ion{Mg}{i} lines relative to other lines 
(\ion{Fe}{i}, \ion{Fe}{ii}, \ion{Ca}{i}, \ion{Ca}{ii}):
$\Delta V_{\scriptscriptstyle \rm Mg\,{\sc II}} = -0.44\pm0.05$ \kms, 
and
$\Delta V_{\scriptscriptstyle \rm Mg\,{\sc I}} = -0.17\pm0.17$ \kms.     
This translates into the isotopic ratio \mga:\mgb:\mgc\ = 
$(19\pm11):(22\pm13):(59\pm6)$ 
with a strong relative overabundance of heavy Mg
isotopes, (\mgb+\mgc)/\mga\ = 4, as compared to the solar ratio 
\mga:\mgb:\mgc\ = 79:10:11, and (\mgb+\mgc)/\mga\ = 0.3.
In the systems at \zabs\ = 1.58 and \zabs\ = 1.65 enriched by AGB-stars we
find only upper limits on the content of heavy Mg isotopes
(\mgb + \mgc)/\mga\ $\la 0.7$ and 
(\mgb + \mgc)/\mga\  $\la 2.6$, respectively. 
At \zabs\ = 1.58, we also put a strong constraint on a putative variation of $\alpha$:
$\Delta \alpha/\alpha = (-1.5 \pm 2.6)\times10^{-6}$ which is one of the most stringent limits
obtained from optical spectra of QSOs. 
We reveal that the wavelength calibration in the range above 7500 \AA\  
is subject to systematic wavelength-dependent drifts. 
}
{} 
\authorrunning{I. I. Agafonova \etal}
\titlerunning{Mg isotope abundances at high redshifts}

\maketitle

\section{Introduction}
\label{sect-1}

Magnesium is one of a few elements for which the isotope abundances can be measured 
from astronomical spectra.
It has three stable isotopes~-- the alpha-nucleus
\mga\ and the neutron-rich nuclei \mgb\ and \mgc.
In the solar photosphere, the magnesium isotopes are mixed in the proportion 
\mga:\mgb:\mgc\ = 78.99:10.00:11.01 (Morton 2003).
High-resolution spectral observations of other stars in the Milky Way show that 
the content of heavy isotopes can be higher, e.g., \mga:\mgb:\mgc\ = 48:13:39  (Yong \etal\ 2003, 2006).
An enhanced content of \mgb, \mgc~-- up to \mga:\mgb:\mgc\ = 60:20:20~-- was also measured in
some presolar spinel grains (Zinner \etal\ 2005; Gyngard \etal\ 2010) 
which are thought to be related to the interstellar dust. 
\mgb\ and \mgc\ are produced from \mga\ via proton capture in the MgAl chain and from $^{22}$Ne via alpha-capture
and are supposed to be enhanced in
ejecta of nova explosions of CO white dwarfs 
where $^{22}$Ne is the third most abundant nuclide (Jos\'e \etal\ 1999, 2004),
or in outflows from AGB-stars where the so-called hot-bottom burning operates (Karakas \etal\ 2003, 2006, 2010).
Direct measurements of Mg isotopic ratio in different objects
allow us to identify the sources of the chemical enrichment and thus  
to understand the chemical evolution of these objects.

Beyond the study of the chemical enrichment of the intergalactic gas,
measurements of the Mg isotopic abundances in the intervening clouds at high redshifts 
are closely related to
the current debate on a putative variation of the fine-structure constant $\alpha$ 
(for a review, see, e.g., Uzan 2010).
Strong resonance lines of the \ion{Mg}{ii} doublet ($\lambda\lambda$2796, 2803 \AA) are often used
as a reference (`anchor') relative to which the
positions of all other ions and, hence, the value of  
$\Delta\alpha/\alpha = (\alpha_{\rm space} - \alpha_{\rm lab})/\alpha_{\rm lab}$
is calculated (Murphy \etal\ 2001; Chand \etal\ 2004).
However, if Mg isotope abundances in quasar absorbers differ from the solar value, this
manifests itself in a shift of the reference lines and, hence, can mimic variations of $\alpha$ 
(Levshakov 1994; Ashenfelter \etal\ 2004). 

In our recent paper (Levshakov \etal\ 2009, hereafter Paper~I)
we reported on a tentative detection
of a blueward shift of the \ion{Mg}{ii} $\lambda\lambda$2796, 2803 \AA\ lines 
in several metal-rich absorbers at \zabs\ $\sim 1.8$,
which may be caused by an enhanced content of 
\mgb\ and \mgc\ in the absorbing gas. 
However, this was a rather qualitative result
since the available quasar spectra had neither a sufficiently high spectral resolution 
nor a calibration accuracy needed to evaluate a particular isotopic abundance ratio.

Heavy Mg isotopes have shorter wavelengths than \mga, and an enhanced content of \mgb\ and \mgc\ in the
mixture \mga+\mgb+\mgc\ can be revealed from a blueward shift of the barycenter of Mg lines relative to
its value based on the laboratory measurements (i.e., with the solar abundance ratio).
The maximum possible isotope shift for the \ion{Mg}{ii} lines is $-0.8$ \kms~-- when all absorbing Mg
consists of \mgc.  For the isotope ratios measured in the Milky Way giants and in presolar spinel grains
the offset of \ion{Mg}{ii} lines would be several times smaller~-- from $-0.25$ to $-0.1$ \kms.
Since the limiting pixel size which can be achieved in spectral observations of
such faint sources as quasars is of $\sim 1$ \kms,
all expected blueward shifts of the \ion{Mg}{ii} lines caused by the presence
of neutron-rich species are at the sub-pixel scale. 
To detect such shifts we need a very bright object to reach a high S/N ratio
($\ga 30$) even at high spectral resolutions. 
Besides, in order to distinguish isotope shifts of the \ion{Mg}{ii} lines
from kinematic shifts caused by the velocity-density inhomogeneities inside the absorber, 
a special attention should be paid for the selection of absorbers: 
appropriate systems ought to contain 
narrow lines of several low-ionization metal ions with simple symmetric profiles.

While performing the measurements on the sub-pixel level, 
the calibration accuracy becomes of the utmost importance. 
Recent tests with iodine cells for both Keck/HIRES and VLT/UVES spectra show that 
calibration uncertainties
can result in wavelength offsets up to 1000 \ms\ 
(Griest \etal\ 2010; Whitmore \etal\ 2010).
The local wavelength calibration can be verified by
comparison of different lines of the same ion like, e.g., 
\ion{Si}{ii} $\lambda\lambda 1260, 1304, 1526$ \AA, or 
\ion{Fe}{ii} $\lambda\lambda 2344, 2382, 2600$ \AA.
This again requires the presence of such absorption systems in quasar spectra. 

Accounting for all these conditions, a
bright quasar \object{HE0001--2340} ($V = 16.6$, $z_{\rm em} = 2.28$) 
discovered in course of the Hamburg/ESO survey
(Wisotzki \etal\ 1996; Reimers \etal\ 1996; Wisotzki \etal\ 2000)
seems to be a perfect target: several systems with strong
and narrow \ion{Mg}{ii} lines
accompanied by lines of other low-ionization ions are identified in its spectrum. 
These are the systems at \zabs\ = 0.45207 (D'Odorico 2007),
\zabs\ = 1.5864 (Chand \etal\ 2004), \zabs\ = 1.6515 (Paper~I) 
which are suitable for measuring the Mg isotope abundances, 
and a sub-DLA system at \zabs\ = 2.1871 (Richter \etal\ 2005)
exhibiting several sets of
absorption lines of the same ion which can be used to test the local wavelength calibration. 

In this paper the subsequent Sect.~\ref{sect-2} presents 
observations and data reduction, the 
analysis of the line profiles is described in Sect.~\ref{sect-3}, 
the results obtained are discussed in Sect.~\ref{sect-4},
and Sect.~\ref{sect-5} sums up the present study.

\section{Observations}
\label{sect-2}

\addtocounter{table}{1}

The observations of the quasar \object{HE0001--2340} 
were acquired with the UV-Visual Echelle 
Spectrograph (UVES) at the VLT 8.2-m telescope at Paranal (Chile) on  
5 nights in September 20-24, 2009\footnote{Program ID 083.A.0733(A)}. The journal of observations
is reported in Table~\ref{tbl-1} together with additional information.
Some exposures were taken with the standard dichroic beam splitter DIC2 (settings 437+760 and 420+700), 
and some with the DIC1 (settings 390+580) thus providing the total wavelength coverage between 350 and 860 nm.
The wavelength ranges covered by individual settings are given in Table~\ref{tbl-2}.
The slit width was
set to 0.7 arcsec for all observations providing 
a resolving power of $\approx 65554\pm3868$. 
This slit width is narrower than usually used for observing high-redshift quasars. 
During the observations the seeing was between  0.5 arcsec to 1 arcsec as
measured by the Differential Image Motion Monitor (Sarazin \& Roddier 1990) 
but generally was slightly better at the telescope.
The major difference with all previous observations was that the CCD pixels
were read without binning.
Data reduction was performed with the last version 4.4.8 
of the UVES pipeline (Larson \&  Modigliani 2009).
The optimal extraction
method was adopted to extract the flux in the cross dispersion direction. 
The 2D images of the 
long-slit calibration lamp and long-slit flat-field have been bias subtracted but not flat fielded
(in the UVES, a pinhole lamp is used for the location of the orders which are curved and somewhat
tilted upward).
In the final spectrum    
the pixel size is of $\approx 1.25$ \kms\ (FWHM $\approx 5$ \kms).

On every night, the calibration spectra were taken immediately before and after the 
scientific exposures. This ensures that any environmental changes occurring during the observations
are accurately accounted for.
Since December 2002, the UVES was equipped with an automatic resetting of the Cross Disperser encoder
positions at the start of each 
exposure\footnote{http://www.eso.org/observing/dfo/quality/UVES/pipeline/\\pipe\_reduc.html}. 
The reason is to use the daytime ThAr calibration frames for saving the night time. This is fully
justified for standard observations, but not for the measurements at the
sub-pixel level which require the best possible wavelength calibration.
To avoid the spectrograph resetting at the start of every exposure, 
the calibration frames were taken in the {\it attach} mode. The wavelength solutions
were determined from each of the ThAr exposures and 
applied to the corresponding quasar exposures to make the pixel-to-wavelength conversion.  
For each frame of about 400 ThAr lines, more than 55\% of the 
lines in the region were used to
calibrate the lamp exposures.  A polynome of the 5th order was adopted.  
Residuals of the wavelength calibrations were typically of  
25 \ms\ and symmetrically distributed around the final
wavelength solution at all wavelengths. 
However, this represents the formal precision of the calibration curve but not
the real calibration accuracy.


\longtab{1}{
\begin{longtable}{ccccllllll}
\caption{\label{tbl-1} Journal of the observations. In the first column, calibration exposures are not numbered 
(blank).
$T_{\rm min}$, $T_{\rm max}$ are minimum and maximum air temperature during the exposure, and
$P_{\rm start}$, $P_{\rm end}$ are air pressure at the beginning and end of the exposure. ENC~-- encoder readings. 
}\\             
\hline\hline       
No. of &Date  & Starting  & Setting, &Exposure, & $T_{\rm min}$, & $T_{\rm max}$, &  
$P_{\rm start}$, &$P_{\rm end}$, &ENC \\ 
exposure & & time, UT &$\lambda$ (nm) &sec & \multicolumn{1}{c}{K} & \multicolumn{1}{c}{K} &  mmHg  & mmHg \\                
\hline       
\noalign{\smallskip}
\endfirsthead
\caption{continued.}\\
\hline\hline       
No. of&Date  & Starting  & Setting, &Exposure, & $T_{\rm min}$, & $T_{\rm max}$, &  
$P_{\rm start}$, &$P_{\rm end}$, &ENC \\ 
exposure & & time, UT &$\lambda$ (nm) &sec & \multicolumn{1}{c}{K} & \multicolumn{1}{c}{K} &  mmHg  & mmHg \\                
\hline       
\noalign{\smallskip}
\endhead
\hline
\endfoot
 &  2009-09-20 & 01:38:16 & 437& 25 & 12.5 & 12.5  &   743.53 & 743.53& 1079359\\[-3pt]
 &  2009-09-20 & 01:38:12 & 760& 1 & 12.4 & 12.5   &   743.53 & 743.53& 557957\\[-3pt]
1 &  2009-09-20 & 01:39:46 & 760& 5400 & 12.4 & 12.5&  743.57 & 743.75& 557957\\[-3pt]
1 &  2009-09-20 & 01:39:50 & 437& 5400 & 12.4 & 12.5&  743.57 & 743.75& 1079359\\[-3pt]
 &  2009-09-20 & 03:11:00 & 437& 25 & 12.4& 12.4&      743.73 & 743.72&1079359\\[-3pt]
2&  2009-09-20 & 03:12:35&760& 5400& 12.4 & 12.4 &     743.77 & 743.63&557954\\[-3pt]
2&  2009-09-20 & 03:12:39&437& 5400& 12.4 & 12.4 &     743.77 & 743.63&1079359\\[-3pt]
 &  2009-09-20 & 04:43:46 & 437& 25 & 12.4& 12.4&      743.60 & 743.62&1079359\\[-3pt]
 &  2009-09-20 & 04:43:43 & 760& 1 & 12.4& 12.4&       743.60 & 743.62&557954\\[-3pt]
 &  2009-09-20& 06:46:15&437& 25& 12.1& 12.2 &         743.03 & 743.03&1079357\\[-3pt]
 3&  2009-09-20& 06:47:36&437& 5400& 11.9& 12.1 &      743.00 & 742.70&1079357\\[-3pt]
 3&  2009-09-20& 06:47:32&760& 5400& 11.9& 12.1 &      743.00 & 742.70&557947\\[-3pt]
 &  2009-09-20& 08:18:43&437& 25& 12.1& 12.2 &         742.68 & 742.68&1079357\\
&  2009-09-21 & 01:44:27 & 580& 25 & 11.7 & 11.7 &     744.13 & 744.18&4690970\\[-3pt]
&  2009-09-21 & 01:44:30 & 390& 25 & 11.7 & 11.7 &     744.13 & 744.18&1093943\\[-3pt]
4&  2009-09-21 & 01:46:01 & 580& 5400 & 11.7 & 11.7 &  744.17 & 743.88&4690970\\[-3pt]
4&  2009-09-21 & 01:46:05 & 390& 5400 & 11.7 & 11.7 &  744.17 & 743.88&1093943\\[-3pt]
&  2009-09-21 & 03:17:09 & 580& 25 & 11.7 & 11.7 &     743.88 & 743.88&4690970\\[-3pt]
&  2009-09-21 & 03:17:12 & 390& 25& 11.7 & 11.7&       743.88 & 743.88& 1093943\\[-3pt]
5&  2009-09-21 & 03:18:44 & 580 & 5400 & 11.6 & 11.7&  743.88 & 743.68& 4690970\\[-3pt]
5&  2009-09-21 & 03:18:48 & 390& 4507 & 11.6 & 11.7&   743.88 & 743.68& 1093943\\[-3pt]
&  2009-09-21 & 04:35:12 & 580& 25 & 11.6 & 11.6 &     743.70 & 743.68&4690970\\[-3pt]
&  2009-09-21 & 04:35:14 & 390& 25& 11.6 & 11.6&       743.70 & 743.68& 1093943\\
&  2009-09-22 & 00:12:33 & 420& 25 & 11.7 & 11.7 &     742.88 & 742.88&1084636\\[-3pt]
6&  2009-09-22 & 00:14:11 & 420& 5400 & 11.1 & 11.7 &  742.90 & 743.15&1084636\\[-3pt]
6&  2009-09-22 & 00:14:07  & 700& 5400 & 11.1 & 11.7 & 742.90 & 743.15&544738\\[-3pt]
&  2009-09-22 & 01:45:20 & 420& 25 & 11.1 & 11.1 &     743.15 & 743.15&1084636\\[-3pt]
7&  2009-09-22 & 01:46:58 & 700& 5400 & 10.8 & 11.1 &  743.17 & 742.78&544727\\[-3pt]
7&  2009-09-22 & 01:47:02 & 420& 5400 & 10.8 & 11.1 &  743.17 & 742.78&1084634\\[-3pt]
&  2009-09-22 & 03:18:12 & 420& 25 & 10.8 & 10.8 &     742.78 & 742.80&1084634\\[-3pt]
8&  2009-09-22 & 03:19:55 & 420& 4279 & 10.8 & 10.8 &  742.78 & 742.48&1084632\\[-3pt]
8&  2009-09-22 & 03:19:51 & 700& 4200 & 10.8 & 10.8 &  742.78 & 742.50&544720\\[-3pt]
&  2009-09-22 & 04:32:19 & 420& 25 & 10.8 & 10.8 &     742.48 & 742.50&1084632\\[-3pt]
 &  2009-09-22 & 04:32:10 & 700& 1 & 10.8 & 10.8 &     742.48 & 742.48&544720\\
&  2009-09-23 & 00:21:28  & 580& 15 & 10.4 & 10.4 &    742.30 & 742.30&4690939\\[-3pt]
&  2009-09-23 & 00:21:31  & 390& 25 & 10.4 & 10.4 &    742.30 & 742.30&1093936\\[-3pt]    
9&  2009-09-23 & 00:22:51  & 580& 5400 & 10.3 & 10.4 & 742.33 & 742.38&4690939\\[-3pt]
9&  2009-09-23 & 00:22:54  & 390& 5400 & 10.3 & 10.4 & 742.33 & 742.38&1093936\\[-3pt]
&  2009-09-23 & 02:03:37  & 390& 25 & 10.3 & 10.3 &    742.38 & 742.40&1093936\\[-3pt]      
&  2009-09-23 & 02:03:34  & 580& 15 & 10.3 & 10.3 &    742.38 & 742.40&4690939\\[-3pt]
10&  2009-09-23 & 02:05:17 & 580 & 3600& 10.2 &10.3 &  742.38 & 742.15& 4690933\\[-3pt] 
10&  2009-09-23 & 02:05:18 & 390  & 3600 &10.2 &10.3 & 742.38 & 742.15& 1093935\\[-3pt]  
&  2009-09-23 & 03:06:27 &  580 & 15  & 10.2 & 10.2 &  742.17 & 742.20& 4690933\\[-3pt]
&  2009-09-23 & 03:06:29 &  390 & 25  & 10.2 & 10.2 &  742.17 & 742.20&1093935 \\[-3pt] 
11&  2009-09-23 & 03:07:50 & 580 & 3600& 10.2 & 10.2 & 742.13 & 741.87&4690933 \\[-3pt]
11&  2009-09-23 & 03:07:52 & 390 & 3864 & 10.2 & 10.2 &742.13 & 741.83& 1093935\\[-3pt]  
&  2009-09-23 & 04:13:13 &  580 & 25  & 10.2 & 10.2 &  741.83 & 741.87&4690933 \\[-3pt]
&  2009-09-23 & 04:13:23 &  390 & 25  & 10.2 & 10.2 &  741.83 & 741.87&1093935\\       
12&  2009-09-24 & 02:14:01 & 580 & 3740 &11.1 & 11.3 & 741.87  & 741.88& 4690957 \\[-3pt]
12&  2009-09-24 & 02:14:05 & 390  & 3740 &11.1 & 11.3 & 741.90  & 741.88& 1093940 \\[-3pt]          
&  2009-09-24& 03:17:29 & 580& 15 & 11.1 & 11.1 &  741.90& 741.88&4690957\\[-3pt]  
&  2009-09-24& 03:17:33& 390 & 15 & 11.1 & 11.1 &  741.90& 741.88& 1093940\\[-3pt]                  
13&  2009-09-24& 03:19:21 & 580& 3740 & 11.1 & 11.1 &  741.88& 741.88&4690954\\[-3pt]  
13&  2009-09-24& 03:19:25 & 390& 3740 & 11.1 & 11.1 &  741.88& 741.88&1093939\\[-3pt]               
&  2009-09-24& 04:22:49 & 580& 15 & 11.1 & 11.1 &  741.90& 741.90&4690954\\[-3pt] 
&  2009-09-24& 04:22:52 & 390& 25 & 11.1 & 11.1 &  741.90& 741.90&1093939\\[-3pt]                   
14& 2009-09-24 & 05:17:51  & 580 & 3740 & 11.0 & 11.1 &741.70& 741.50&4690952\\[-3pt]
14& 2009-09-24 & 05:17:55  & 390  & 3740 & 11.0 & 11.1 &741.70& 741.50&1093939\\[-3pt]              
& 2009-09-24 & 06:21:19  & 580 & 15& 11.1 & 11.1 &741.50& 741.48&4690952\\[-3pt]
& 2009-09-24 & 06:21:22  & 390 & 15& 11.0 & 11.1 &741.50& 741.48&1093939\\ 
\end{longtable}
} 

\begin{table}[t!]
\centering
\caption{Wavelength ranges covered by individual settings}
\label{tbl-2}
\begin{tabular}{lc}
\hline
\hline
\noalign{\smallskip}
Setting & $\Delta \lambda$, \AA \\
\noalign{\smallskip}
\hline
390 & $3290-4519$\\
420 & $3586-4820$\\
437 & $3759-4985$\\
580l & $4788-5763$\\
580u & $5836-6809$\\
700l & $5530-6936$\\
700u & $7050-8904$\\
760l & $5694-7532$\\
760u & $7660-9465$\\
\noalign{\smallskip}
\hline
\end{tabular}
\end{table}

The precision of the wavelength calibration in the course of science reduction is determined by 
the temperature and air pressure gradients between the times of the science 
exposure and of the wavelength calibration applied to the data since the 
environmental changes may cause drifts in the
refractive index of air inside the spectrograph between the ThAr and quasar exposures. 
These thermal-pressure drifts move the different cross dispersers in different ways
thus introducing the relative shifts between the different spectral ranges for 
different exposures. 
For instance, 
according to Kaufer \etal\ (2004) the pressure changes in the UVES enclosure 
induce a shift of $\approx 66.7$ \ms/mmHg, 
whereas temperature fluctuations produce shifts of $\approx 420$ \ms/degr~C in the red arm and 
$\approx 60$ \ms/degr~C in the blue 
arm\footnote{http://www.eso.org/observing/dfo/quality/UVES/pipeline/\\pipe\_reduc.html}.
The magnitudes of the possible shifts in our dataset can be estimated from 
the minimum and maximum temperatures inside the spectrograph enclosure
during the exposures and from the pressure values at the beginning and the end of each exposure
which are given in Table~\ref{tbl-1}.
In the last column of this table the encoder readings are reported which indicate that there were 
no grating resets within each block of observations. We emphasize that this effect
has never been taken into account in all the studies performed so far on UVES data. 
There are no measurable temperature changes for the short exposures of the calibration lamps
but during the science exposures the temperature drifted generally by 0.1-0.2 K.  
Pressure changes range from 0.15 to 0.6 mmHg (note that the start and the end pressure values
are inverted in the UVES fits headers).   
Thus, measured
temperature and pressure changes assure a radial velocity stability within $\approx$ 50 \ms\
in the blue arm and within $\approx$ 100 \ms\ in the red arm.

Individual spectra were corrected for the motion of the observatory around the barycenter of the
Earth-Sun system and reduced to vacuum. The component of the observatory's
barycentric velocity in the direction to the object 
was calculated using the date and time of the integration midpoint.
The air wavelengths were
transformed to vacuum by means of the dispersion formula by  
Edl\'en (1966). 
We note that the barycentric and vacuum corrections require a rebinning which
introduces a certain degree of correlations between the fluxes in the
adjacent pixels.

\begin{table}[t!]
\centering
\caption{Atomic data for \ion{Mg}{i}, \ion{Mg}{ii},  \ion{Al}{ii}, \ion{Al}{iii}, \ion{Si}{ii}, 
\ion{Ca}{i}, \ion{Ca}{ii}, \ion{Fe}{i}, and \ion{Fe}{ii} resonance transitions}
\label{tbl-3}
\begin{tabular}{l r@{.}l r@{.}l c}
\hline
\hline
\noalign{\smallskip}
Line & \multicolumn{2}{c}{$\lambda^\ast_{\rm vac}$, \AA } & \multicolumn{2}{c}{$f^\dagger$} & $\lambda_{\rm vac}$ Ref \\
\noalign{\smallskip}
\hline
\noalign{\smallskip}
Mg\,{\sc i} & 2852&96282(10)& 1&83 & [1] \\[2pt]
Mg\,{\sc ii} & 2803&53112(10)& 0&306 & [1] \\[-1pt]
Mg\,{\sc ii} & 2796&35403(10)& 0&615 & [1]  \\[2pt]
Al\,{\sc ii} & 1670&7886& 1&74 & [2] \\[2pt]
Al\,{\sc iii} & 1862&7910& 0&278 & [2] \\[-1pt]
Al\,{\sc iii} & 1854&7184& 0&542 & [2] \\[2pt]
Si\,{\sc ii} & 1526&7070& 0&133 & [2] \\[-1pt]
Si\,{\sc ii} & 1304&3702& 0&0863 & [2] \\[-1pt]
Si\,{\sc ii} & 1260&4221& 1&18 & [2] \\[2pt]
Ca\,{\sc i} & 4227&918& 1&77 & [2] \\[2pt]
Ca\,{\sc ii} & 3969&5901& 0&3116 & [2] \\[-1pt]
Ca\,{\sc ii} & 3934&7750& 0&6267 & [2] \\[2pt]
Fe\,{\sc i} & 3720&9928 & 0&0413  & [2] \\[-1pt]
Fe\,{\sc i} & 3021&5187 & 0&104   & [2] \\[-1pt]
Fe\,{\sc i} & 2967&7646 & 0&0438  & [2] \\[-1pt]
Fe\,{\sc i} & 2523&6083 & 0&198   & [2] \\[-1pt]
Fe\,{\sc i} & 2484&0209 & 0&520   & [2] \\[2pt]
Fe\,{\sc ii} & 2600&17223(10) & 0&239  & [1] \\[-1pt]
Fe\,{\sc ii} & 2586&64937(10) & 0&0691 & [1]  \\[-1pt]
Fe\,{\sc ii} & 2382&76411(10) & 0&320  & [1]  \\[-1pt]
Fe\,{\sc ii} & 2374&46013(10) & 0&0313 & [1]  \\[-1pt]
Fe\,{\sc ii} & 2344&21282(10) & 0&114  & [1]  \\[-1pt]
Fe\,{\sc ii} & 1608&45078(5)& 0&0577   & [3] \\[2pt]
\noalign{\smallskip}
\hline
\noalign{\smallskip}
\multicolumn{6}{l}{\footnotesize References: [1]~-- Aldenius (2009); [2]~-- Morton (2003);}\\
\multicolumn{6}{l}{\footnotesize [3]~-- Nave \& Sansonetti (2010).}\\
\multicolumn{6}{l}{\footnotesize $^\ast$$\lambda_{\rm vac}$ absolute errors 
affect the second to last digit in [2].}\\
\multicolumn{6}{l}{\footnotesize $^\dagger$Oscillator strengths $f$ are taken from [2].}\\
\end{tabular}
\end{table}

\section{Analysis}
\label{sect-3}

The laboratory line wavelengths used throughout the paper are listed in Table~\ref{tbl-3}.
For \ion{Mg}{i}, \ion{Mg}{ii}, and \ion{Fe}{ii} lines (except for $\lambda1608$ \AA),
we adopted the wavelengths from Aldenius (2009).
The wavelengths of the \ion{Mg}{i} line from Salumbides \etal\ (2006), 
the \ion{Mg}{ii} lines from Batteiger \etal\ (2009),
and the \ion{Fe}{ii} lines from Nave \& Sansonetti (2010) all are shorter by $\la 0.2$ m\AA\ than
the corresponding values from Aldenius (2009). 
In the velocity space, the shifts between these wavelength scales are in the range from
3.5 \ms\ to 26 \ms, but since all wavelengths are shifted in the same direction, the relative
shifts between lines which are compared in our analysis do not exceed 15 \ms\ 
which is well below the measurement errors. 
The photospheric solar abundances are taken from Lodders \etal\ (2009).

The ThAr wavelength calibration of QSO spectra taken with slit echelle
spectrographs such as, e.g., the Keck/HIRES or the VLT/UVES, is not stable and can
produce both velocity offsets between different exposures and intra-order velocity distortions 
within each exposure (Levshakov \etal\ 2006; Centuri\'on \etal\ 2009; Griest \etal\ 2010;
Whitmore \etal\ 2010). 
The reason of these calibration distortions is still not well understood. 
They can be caused by differences in the slit illumination between the source and the calibration lamp,
by non-linearity of echelle orders, non-uniformity in the distribution of the reference ThAr lines, and by
other still obscure factors. In general, the detected velocity shifts are of order of a few hundreds of \ms, i.e.
well below the pixel size. 
Thus, if the effects to be studied have a characteristic scale of about 
the pixel size, QSO spectra can be prepared by means of common pipeline procedures. 
In our case the pixel size is $\simeq 1.3$ \kms\
and the expected line shifts are about hundreds of \ms, i.e., 
we work on a sub-pixel level.
In order to guarantee the required accuracy, the profiles of all absorption lines used in the present study
were prepared individually in accord with the following procedure.

\begin{figure}[t]
\vspace{0.0cm}
\hspace{-3.0cm}\psfig{figure=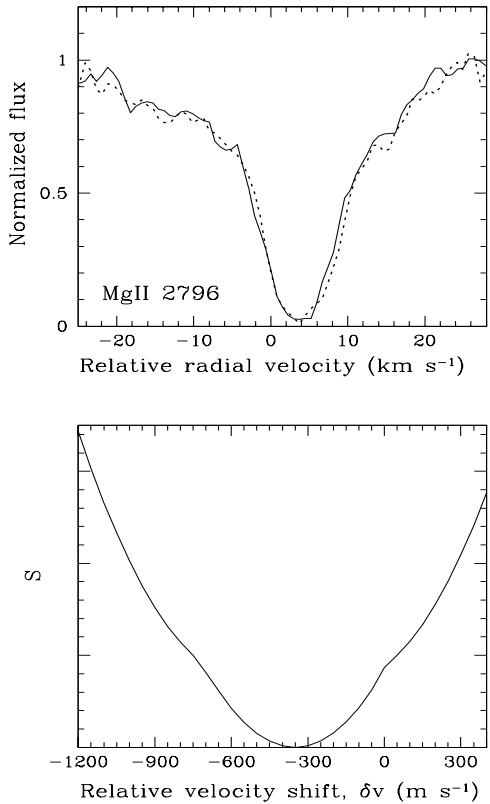,height=14cm,width=16cm}
\vspace{-3.5cm}
\caption[]{An example of the cross-correlation of two profiles from
different exposures.
{\it Upper panel:} Normalized intensities 
of the \ion{Mg}{ii} $\lambda2796$ \AA\ line from the \zabs\ = 1.5864 system
(Sect.~\ref{sect-4-2})
extracted from
exposure \#2, setting 760 (solid line) and from 
exposure \#1, setting 760 (dotted line). See Table~\ref{tbl-1}.
{\it Lower panel:} The dependence of ${\cal S}$~-- the sum of squares of the intensity
differences over all pixels covering the line profiles~-- as function of the
velocity shift, $\delta v$. The profiles become fully aligned when
the dotted line profile is shifted by $\approx -350$ \ms\
($\approx 1/4$ pixel size).
}
\label{fg1}
\end{figure}

At first, we calculate the redshift \zabs\ of a given absorption-line system. 
This redshift is 
fixed for all other lines from this system. 
All line position measurements are carried out in the velocity scale.
Namely, for each absorption line with the laboratory wavelength $\lambda_0$, 
a spectral region centered at $\lambda_{\rm obs} = \lambda_0\ (1 + z_{\rm abs})$ and wide
enough to incorporate clear `continuum windows' from both sides (blueward and redward) of $\lambda_{\rm obs}$ 
is converted into the velocity scale as $V = c (\lambda - \lambda_{\rm obs})/\lambda_{\rm obs}$, where
$c$ is the speed of light and $\lambda$ is the current wavelength. 
The continuum windows are used to determine the local continuum level 
and to normalize the extracted line profiles.

As already mentioned above, the profiles from individual exposures may be shifted relative to each other.
To evaluate these shifts 
we used a procedure that cross-correlates two normalized profiles $I_1$ and $I_2$: 
one of the chosen profiles ($I_1$) is fixed and the other ($I_2$) 
is moved relative to $I_1$ by a small step, $\delta v$
(usually $\delta v \approx 50$ \ms, i.e., of about 0.05 pixel size). 
At each step, a value of {$\cal{S}$}~-- the sum of squares of the intensity
differences over $n$ pixels covering the line~-- is calculated, 
${\cal{S}}(\delta v) = \sum_{j=1}^{n} [I_{1,j} - I_{2,j}(\delta v)]^2$.
{$\cal{S}$} depends parabolically on $\delta v$, being
minimal when the profiles are aligned. This shift at the minimum of ${\cal{S}}$ is taken as
a local velocity offset, $\Delta V$, between the two profiles. 
An example of such cross-correlation is shown in Fig.~\ref{fg1}.

Cross-correlating the profiles of different lines from different exposures 
we found that ($i$) the velocity shifts between individual exposures of the same line 
were, in general, within $|\Delta V| \leq 500$ \ms, 
($ii$) the shift between two arbitrary chosen
exposures was not a constant but varied from line to line (i.e., wavelength dependent).
This supports the above statement that calibration accuracy is affected by many factors. 
Then, a natural solution to obtain
the resulting line profile would be to average over all factors, i.e., to sum up all available exposures. 

However, a particular feature of the present observations is that 
because of the seeing conditions the signal changes
significantly between exposures of different settings. 
Namely, for the setting 390+580, 8 exposures were taken, all with a 
comparable signal level. 
But for the setting 437+760, 
only 3 exposures were obtained, each of which with a signal 
much higher than in the setting 390+580. 
The high-signal exposures 
have larger weights in the sum
and,
hence, even a single miscalibrated high-signal profile can noticeably shift the combined profile.

\begin{figure*}[t]
\vspace{0.0cm}
\hspace{0.0cm}\psfig{figure=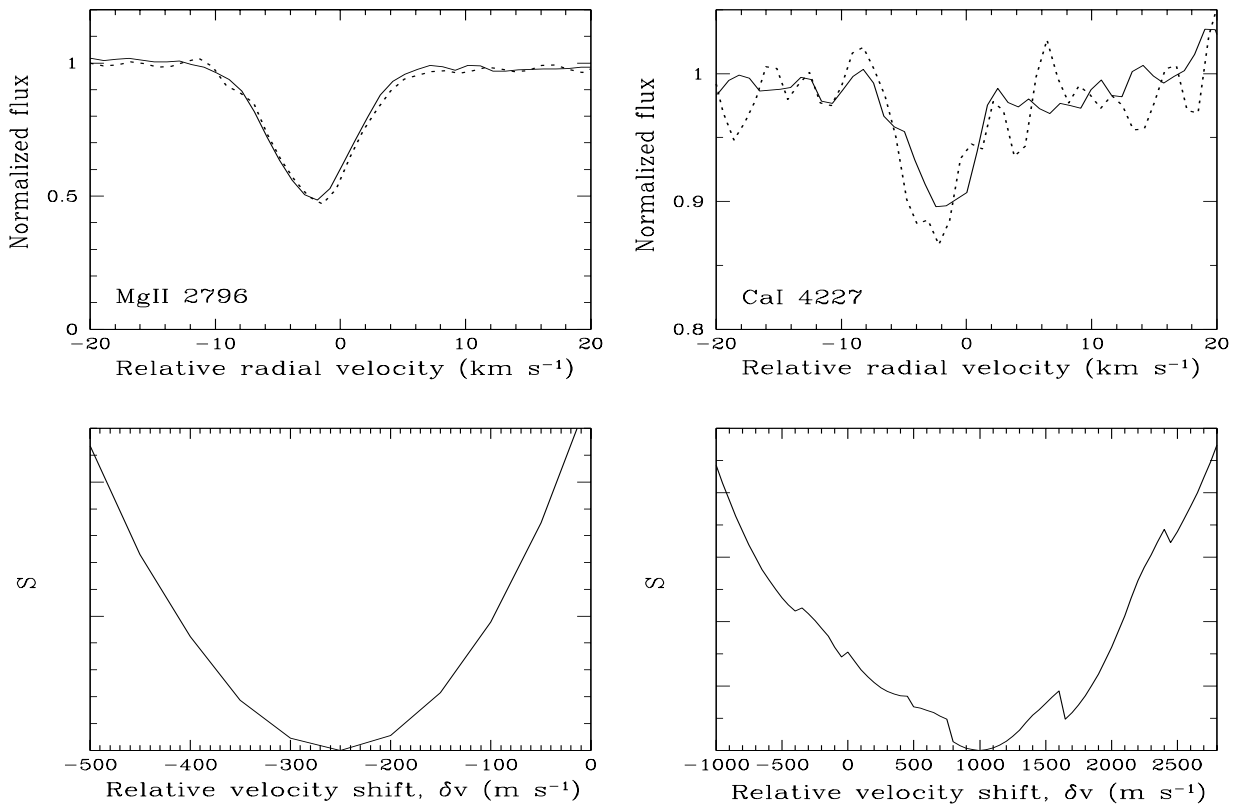,height=14cm,width=18cm}
\vspace{-3.0cm}
\caption[]{Cross-correlations of profiles from exposures with high signal levels.
All lines are from the \zabs\ = 0.45207 system. 
{\it Left upper panel:} Solid line~-- the normalized \ion{Mg}{ii} $\lambda2796$ \AA\
profile composed from 12 exposures with similar signal levels ($100-150$ counts per pixel at the continuum level);
dotted line~-- the normalized \ion{Mg}{ii} $\lambda2796$ \AA\ profile from the exposure \#2,
setting 437
(Table~\ref{tbl-1}) with a high level of signal ($\sim 500$ counts per pixel).
{\it Left lower panel:} ${\cal S}$ as function of $\delta v$ (same as in Fig.~\ref{fg1}).
The dotted line profile is to be shifted by $\approx -250$ \ms\ 
to become fully aligned with the composite profile. 
{\it Right upper panel:} Solid line~-- the normalized profile of the weak \ion{Ca}{i} $\lambda4227$ \AA\
line composed from 8 exposures with similar signal ($\sim 300$ counts per pixel);
dotted line~-- the normalized \ion{Ca}{i} $\lambda4227$ \AA\ profile from the exposure \#1, setting 760
(Table~\ref{tbl-1}) with a high level of signal ($\sim 1000$ counts per pixel).
{\it Right lower panel:} ${\cal S}$ as function of $\delta v$ (same as in Fig.~\ref{fg1}).
The dotted line profile is distorted by noise which leads to its apparent shift relative to
the composite profile of about $-1000$ \ms. The real shift of this portion of the exposure \#1
is only $-150$ \ms\ as calculated from profiles of a strong \ion{Fe}{ii} 
$\lambda2382$ \AA\ line at \zabs\ = 1.5864 located in the vicinity of \ion{Ca}{i} $\lambda4227$ \AA.
}
\label{fg2}
\end{figure*}

To avoid such shifts 
we firstly co-added only profiles with a comparable signal and then
checked whether the profiles from the high signal exposures~-- if present~-- 
were shifted with respect to this combined profile or not. 
An example is shown in Fig.~\ref{fg2}, left panel. 
The shifted exposures were aligned with the combined profile and then all available profiles were 
co-added together to provide the final line. 

Calculations showed that the alignment of the high signal exposures resulted in a shift
of the final line center of about 40--50 \ms\ as compared to a line obtained by a simple
co-adding (i.e., without alignment) of all profiles. 
This value is of the same order as the statistical errors with which the centers of strong and
narrow  absorption lines can be measured (see, e.g., Tables~\ref{tbl-4}--\ref{tbl-5})
and, therefore, the alignment is important while dealing with such lines.
On the other hand, the centers of weak lines are measured with errors of 100--200 \ms\
which means that the alignment of the high signal exposures does not lead to statistically
significant differences. 
Nevertheless, in the present study the profiles of weak lines were also prepared with the
alignment since in many cases this operation allowed us to obtain more regular
(smooth) profiles. 
However, the profiles of weak lines can be severely distorted by noise
which would result in artificially exaggerated $\Delta V$ values (Fig.~\ref{fg2}, right panel).  
That is why the exposure shifts were calculated for a strong absorption line located in the vicinity
(within $15-20$ \AA) of a weak line and then applied to the individual profiles of this weak line.

The wavelength range above 7200 \AA\ is covered by the 3 exposures of the setting 760 (divided into low and upper
frames) and 
by the 3 exposures of the setting 700 (also low and upper frames), all having significantly different signals.
For absorption lines from this range, the individual profiles were aligned with a profile from the  exposure
with the highest signal.
In turn, the local calibration (i.e., at the position of a particular line) 
of this exposure was checked using different transitions of the same ion
(see Sect.~\ref{sect-4-2} and \ref{sect-4-4}). 

The combined profiles were normalized by means of
a smoothing spline and then the rms noise was calculated using the continuum windows.
These final profiles were fitted to Gaussians defined by the standard parameters: the component
center $V$, the broadening Doppler parameter $b$, and the column density $N$.
The FWHM values to convolve the synthetic profiles with the spectrograph point-spread function
were estimated from the ThAr lines in the vicinity of particular absorption lines. 
These values range between 4.3 and 5.4 \kms. However, in some cases a simultaneous fitting of several lines of the
same ion shows that the FWHM should be slightly ($\sim5$\%) smaller than the value estimated
from the ThAr lines. 
This can be explained by additional broadening of the ThAr lines caused by the pressure effects 
since these lines arise in dense plasma.
 
The errors of the fitting parameters were calculated via the inverse second derivatives matrix. 
Since we are especially interested in
the accurate error of the line center $V$, this error was additionally evaluated using the
$\Delta \chi^2$-method (Press  \etal\ 1992). 
As the actual error of the line center the larger estimate was accepted.

\begin{figure}[t]
\vspace{-0.5cm}
\hspace{-2.0cm}\psfig{figure=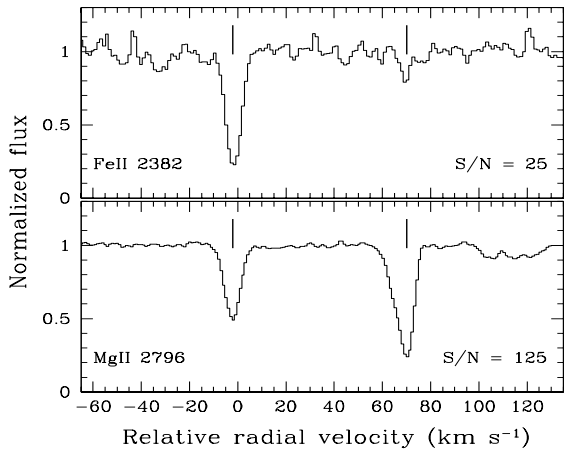,height=14cm,width=16cm}
\vspace{-7.0cm}
\caption[]{
The observed profiles of the Fe\,{\sc II} and Mg\,{\sc II} lines from the \zabs\ = 0.45207 system
towards the quasar \object{HE0001--2340}.
The system consists of two subcomponents separated by $\approx 70$ \kms. 
The zero radial velocity is fixed at $z = 0.45207$.
Note the difference between relative line intensities in the subcomponents.
}
\label{fg3}
\end{figure}

\begin{figure}[t]
\vspace{0.0cm}
\hspace{-1.0cm}\psfig{figure=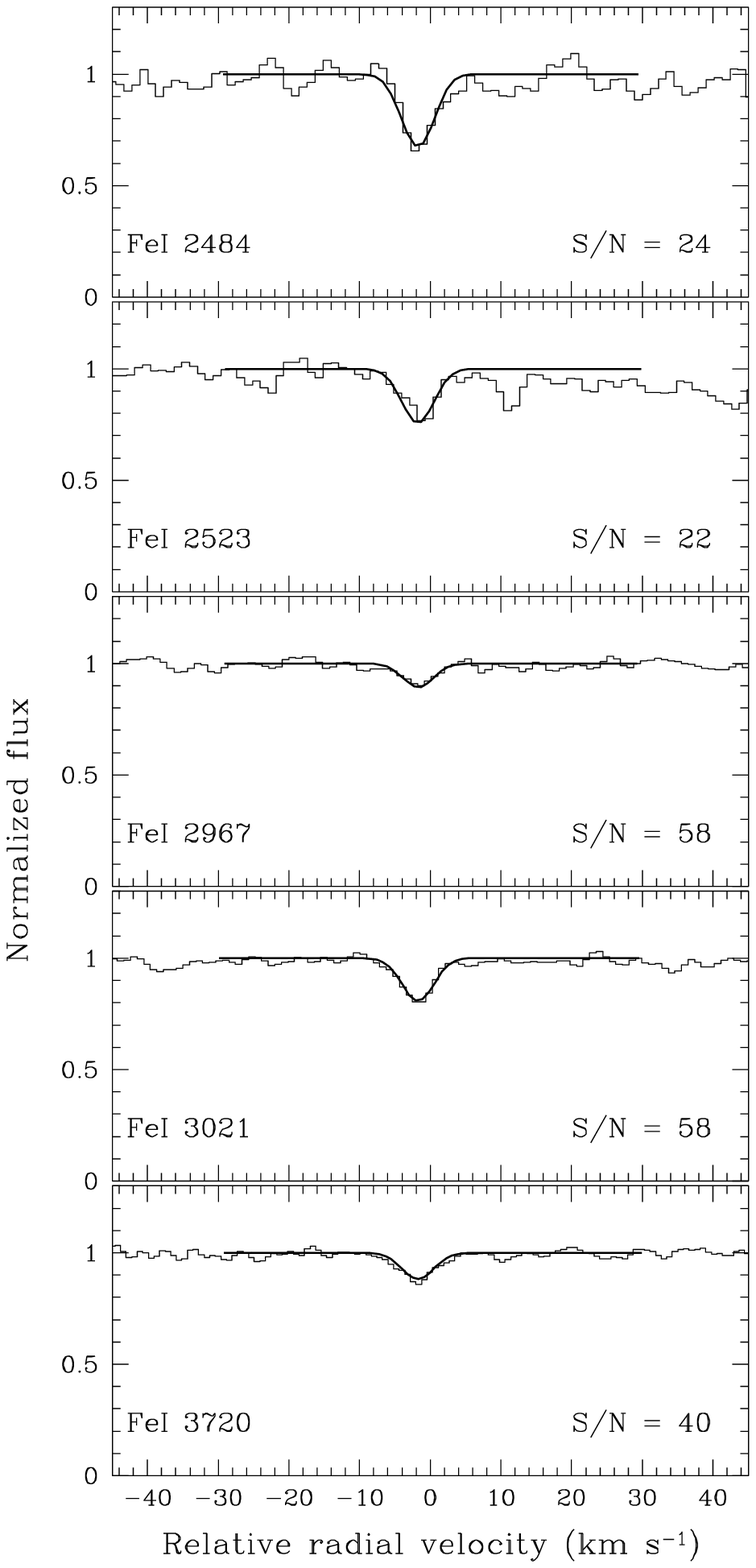,height=14cm,width=16cm}
\vspace{-0.5cm}
\caption[]{
Fe\,{\sc I} lines from the \zabs = 0.45207 system towards the quasar \object{HE0001--2340}
(solid-line histograms).
Synthetic profiles are plotted by the smooth curves.
The signal-to-noise ratio, S/N, per pixel in the co-added spectrum is indicated in each panel.
The zero radial velocity is fixed at $z = 0.45207$.
}
\label{fg4}
\end{figure}

It is well known that the noise produced by CCDs is highly correlated (e.g., Levshakov \etal\ 2002).
These correlations do not affect the results of the $\chi^2$-minimization, but they
reduce the apparent dispersion of the noise and, hence, lead to the underestimation of the
calculated errors of the fitting parameters. 
In order to understand how to correct the estimated errors,
the following Monte Carlo experiment was performed. 
First, we synthesized a line profile
with a given set of parameters. The spectrum of \object{HE0001--2340} reveals wide
continuum windows (e.g., in the range 4474--4692 \AA\ or 5130--5314 \AA) which in fact
represent long sequences (several thousands of points) of correlated noise values. In these
windows, we choose at random a point and starting from it extracted  $n$ subsequent
values which were
added to the synthesized profile ($n$ is the number of pixels covering the line). 
Then the profile parameters were evaluated via the standard $\chi^2$-minimization. 
The procedure was repeated 1000 times, and the dispersions of the resulting distributions
of the fitting parameters were taken as their errors. They were  $\approx 1.5$ times larger than the
errors estimated via the inverse matrix (or the errors estimated from
the  $\Delta \chi^2$-procedure with the confidence level of $\Delta \chi^2 = 1$). Thus, the final
errors listed in Tables~\ref{tbl-4}--\ref{tbl-6} 
below represent the calculated values multiplied by 1.5. 

The difference between the radial velocities $V_1, V_2$ of two absorption lines can be used to estimate a putative
variation of the fine-structure constant $\alpha$ in space and time with respect to its laboratory
(terrestrial) value. For two transitions with different sensitivities to $\alpha$ changes the value of
$\Delta \alpha/\alpha = (\alpha_z - \alpha_{lab})/\alpha_{lab}$ is calculated as
$\Delta \alpha/\alpha = (V_1-V_2)/[2c(Q_2-Q_1)]$, where $c$ is the speed of
light, and $Q_1, Q_2$ are the corresponding sensitivity coefficients 
(Dzuba \etal\ 2002; Berengut \etal\ 2010).

\begin{figure*}[t]
\vspace{0.0cm}
\hspace{0.0cm}\psfig{figure=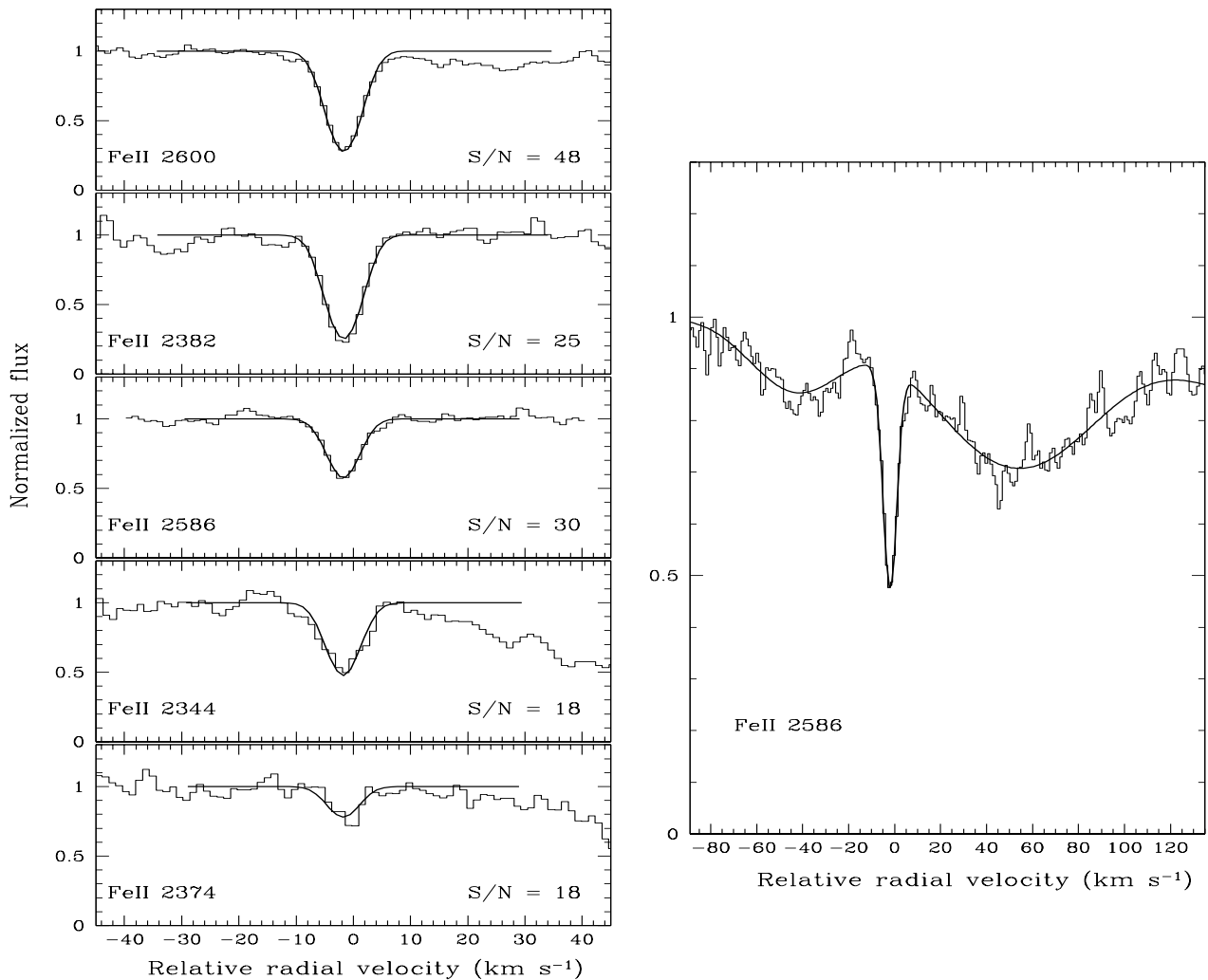,height=14cm,width=18cm}
\vspace{0.0cm}
\caption[]{Same as Fig.~\ref{fg4} but for Fe\,{\sc II} lines. 
The right panel shows the decomposition of the $\lambda2586$ \AA\ line
which is blended with two Ly-$\alpha$ forest absorption lines (see text, for detail).
}
\label{fg5}
\end{figure*}

\begin{figure*}[t]
\vspace{0.0cm}
\hspace{0.0cm}\psfig{figure=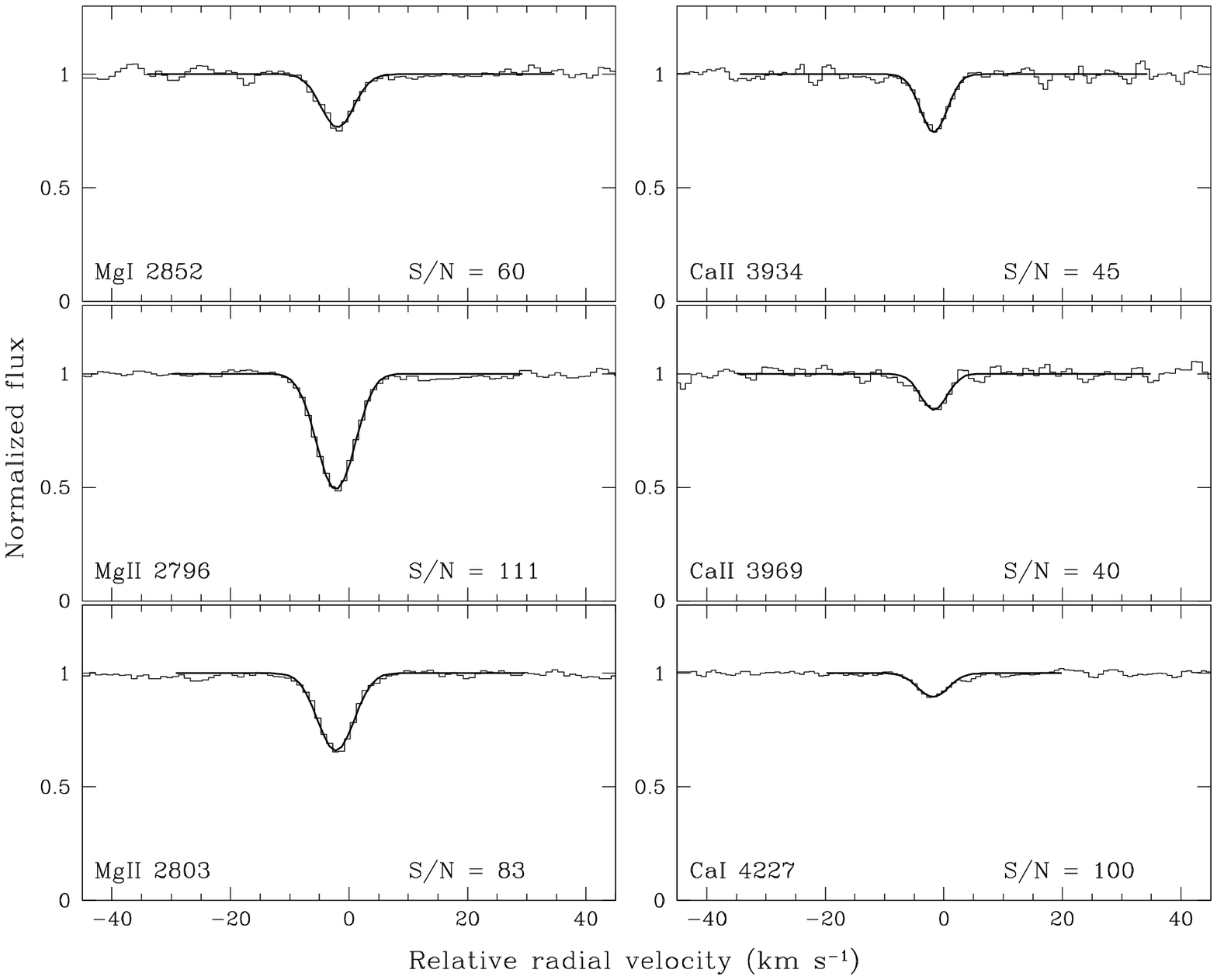,height=14cm,width=18cm}
\vspace{-3.5cm}
\caption[]{Same as Fig.~\ref{fg4} but for Mg\,{\sc I}, Mg\,{\sc II},
Ca\,{\sc I}, and Ca\,{\sc II} lines. 
}
\label{fg6}
\end{figure*}

\section{Results}
\label{sect-4}

\subsection{Absorber at \zabs\ = 0.45207}
\label{sect-4-1}

This unique systems exhibiting a set of \ion{Fe}{i} lines was discovered and firstly described by
D'Odorico (2007).
The system consists of two components separated by $\approx 70$ \kms\ (Fig.~\ref{fg3}). 
The red component shows only a strong \ion{Mg}{ii} doublet and weak \ion{Fe}{ii}
$\lambda\lambda2600, 2382$ \AA\ lines, but the blue component contains, 
along with \ion{Fe}{i}, narrow and well distinguished lines of
\ion{Mg}{ii} $\lambda\lambda2796, 2803$ \AA, 
\ion{Mg}{i} $\lambda2852$ \AA, \ion{Fe}{ii} $\lambda\lambda2600, 2382, 2344, 2374$ \AA, 
\ion{Ca}{ii} $\lambda\lambda3969, 3934$ \AA, and \ion{Ca}{i} $\lambda4227$ \AA\ with simple profiles
and represents in fact the best-choice system to measure the Mg isotope abundance ratio.
In the previous spectrum of \object{HE0001--2340}  taken
with a lower spectral resolution (pixel size $\approx 2.3$ \kms, FWHM = 6.8 \kms), 
the Mn triplet $\lambda\lambda2576, 2594, 2606$ \AA\
can be safely identified and probably also a line of \ion{Si}{i} $\lambda2515$ \AA. 
Unfortunately, in our spectrum which was obtained with
a higher resolution and because of this has a lower S/N these lines are indistinguishable from noise fluctuations.

Calculations were performed with the line profiles prepared as described in Sect.~\ref{sect-3}. 
Some special cases are outlined below.

\ion{Fe}{i} $\lambda2967$ \AA\ (Fig.~\ref{fg4}). The line is weak and not seen in individual exposures. 
However, this \ion{Fe}{i} transition has the highest sensitivity to changes in $\alpha$
(Dzuba \& Flambaum 2008), so it was important to obtain its line profile as accurate as possible.
Fortunately, 12 \AA\ redward to the position of \ion{Fe}{i} $\lambda2967$ 
(at $\lambda \approx 4309$ \AA)  lies a strong 
\ion{Al}{ii} $\lambda1670$ \AA\ line from the \zabs\ = 1.5864 system (Sect.~\ref{sect-4-2}) which can
be used to calculate the shifts of the individual exposures and to test the local wavelength scale calibration. 
The corresponding spectral ranges are present in all 14 exposures~-- 
the 8 exposures of the setting 390 with similar signal,
the 3 high-signal exposures of the setting 437, and the 3 intermediate-signal exposures of the setting 420. 
Employing our standard procedure of the line profile preparation,
we would sum up the 8 equal-signal profiles from the setting 390,
align the profiles from the high-signal exposures with this combined profile,
and then co-add all available exposures. 
I.e., the reference for the final profile  would be set by the average profile from 
the setting 390.
But the comparison of the \ion{Al}{ii} $\lambda1670$ \AA\ line produced by 
the 8 co-added exposures of this setting  
with other low-ionization lines \ion{Si}{ii} $\lambda 1526$ \AA and 
\ion{Fe}{ii} $\lambda\lambda2382, 2600, 2586, 2344, 2374, 1608$ \AA\ from 
the same \zabs\ = 1.5864 system showed that 
the \ion{Al}{ii} profile was
shifted relative to these lines by $\approx 0.5$ \kms, whereas the  profile composed from the combined 6 
exposures of the settings 437 and 420 was coherent with them.
Therefore, in the present case, the reference profile
was obtained by co-adding the exposures from settings 437 and 420,
and the alignment was performed for the individual \ion{Al}{ii} profiles from the setting 390. 
The same procedure was used to prepare
the resulting profile of the \ion{Fe}{i} $\lambda2967$ \AA\ line: 
the exposures of the settings 437 and 420 remained unchanged, whereas those 
from the setting 390 were shifted in accord with 
the shifts calculated for the \ion{Al}{ii} $\lambda1670$ \AA\ profiles. 
Additionally, exposures exhibiting noise spikes within
the \ion{Fe}{i} $\lambda2967$ \AA\
profile were excluded from the final co-adding (3 exposures from the setting 390 and 3 from the setting 420).

\begin{table*}[t!]
\centering
\caption{Fitting parameters of spectral lines identified in the \zabs\ = 0.45207 absorber.
Indicated are $1\sigma$ statistical errors. 
} 
\label{tbl-4}
\begin{tabular}{lcccl}
\hline
\hline
\noalign{\smallskip}
\multicolumn{1}{c}{Ion} & \multicolumn{1}{c}{$V$, \kms} & $b$, \kms & 
\multicolumn{1}{c}{$N$, \cm} & \multicolumn{1}{c}{${\cal C}$} \\
\noalign{\smallskip}
\hline
\noalign{\smallskip}
Mg\,{\sc ii} & $-2.21\pm0.04$ &  $2.79\pm0.25$ & $(2.0\pm0.1)$E12 & 0.77 \\
Mg\,{\sc i}  & $-1.94\pm0.17$ & $2.43\pm0.39$ & $(1.9\pm0.1)$E11 & 0.77$^\ast$ \\
Fe\,{\sc ii} & $-1.77\pm0.06$ & $2.59\pm0.05$ & $(1.06\pm0.03)$E13 & 0.86 \\
Fe\,{\sc i} & $-1.72\pm0.14$ & $0.84\pm0.10$  & $(2.9\pm0.2)$E12 & 0.71  \\
Ca\,{\sc ii} & $-1.82\pm0.16$ & $1.52\pm0.19$ & $(6.7\pm0.4)$E11 & 0.56 \\
Ca\,{\sc i} & $-1.84\pm0.27$ & $1.6\pm1.0$  & $(7.2\pm1.8)$E10 & 0.56$^\ast$\\
\noalign{\smallskip}
\hline
\noalign{\smallskip}
\multicolumn{5}{l}{\footnotesize $^\ast$Single line, covering factor ${\cal C}$ 
adopted from the correspondin}\\[-2pt] 
\multicolumn{5}{l}{ionized doublet.}
\end{tabular}
\end{table*}

\ion{Fe}{i} $\lambda3021$ \AA\ (Fig.~\ref{fg4}). In spite of its weakness, the line is well distinguishable in all
available 14 exposures (settings 390, 437, 420). 
However, from unknown reasons (may be because of the absence of appropriate ThAr lines in the 
vicinity of $\lambda = 4387$ \AA\ where \ion{Fe}{i} $\lambda3021$ is located) the final
profile appeared to be offset
relative to the other \ion{Fe}{i} lines.
However, the sensitivities to $\alpha$ variations
of the \ion{Fe}{i} $\lambda3021$ \AA\ and $\lambda2967$ \AA\ transitions are equal, 
and these lines should be aligned at any rate. 
The cross-correlation of the \ion{Fe}{i} $\lambda3021$ \AA\ and $\lambda2967$ \AA\ lines 
showed that the former was shifted by $-0.35$ \kms. The
calibration of the \ion{Fe}{i} $\lambda2967$ \AA\ line
was verified by the \ion{Al}{ii} $\lambda1670$ \AA\ line from the \zabs\ = 1.5867 system, 
and therefore we aligned the \ion{Fe}{i} $\lambda3021$ \AA\ line 
with \ion{Fe}{i} $\lambda2967$ \AA, i.e., shifted it by 0.35 \kms.

\begin{figure*}[t]
\vspace{0.0cm}
\hspace{0.0cm}\psfig{figure=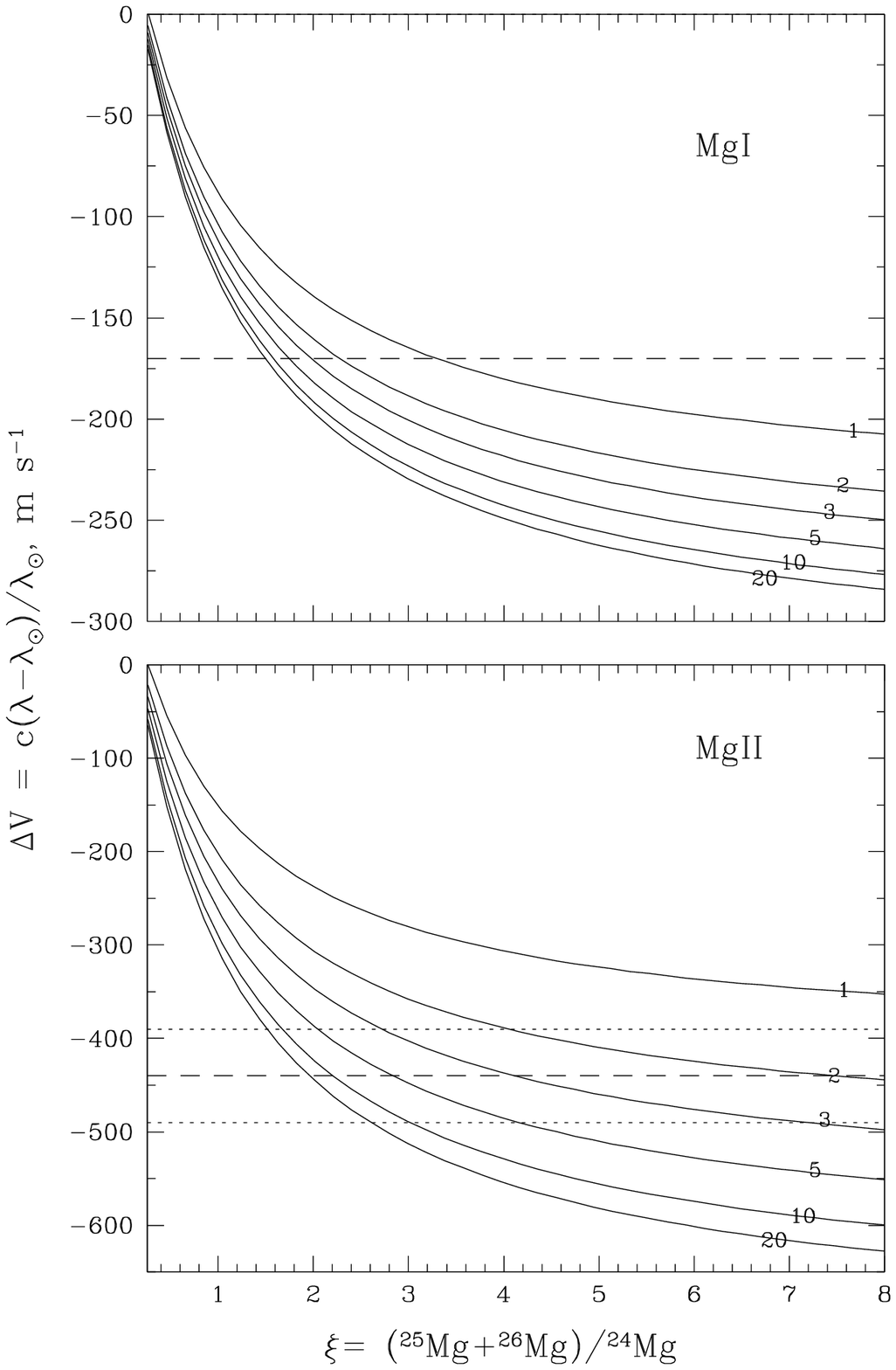,height=18cm,width=18cm}
\vspace{0.0cm}
\caption[]{
The velocity shift $\Delta V$ as function of the ratio  
$\xi =\ $(\mgb + \mgc)/\mga\
for the \ion{Mg}{i} $\lambda2852$ \AA\ (upper panel) and
the \ion{Mg}{ii} $\lambda2796$ \AA\ (lower panel) lines.
The curves are calculated for different ratios
of the heavy isotopes, $r =\ $\mgc/\mgb\ (depicted over each curve).
$\Delta V$ = 0  corresponds to the solar value of the Mg isotope
ratios, $\xi_\odot = 0.2660$ and $r_\odot \approx 1$.  
The horizontal dashed lines mark the mean values of 
$\Delta V_{\scriptscriptstyle \rm Mg\,{\sc I}}$
and $\Delta V_{\scriptscriptstyle \rm Mg\,{\sc II}}$ measured at \zabs = 0.45207, 
whereas the dotted lines restrict the $\pm1\sigma$
uncertainty intervals (see text, for detail).
}
\label{fg7}
\end{figure*}

\ion{Fe}{i} $\lambda3720$ \AA\ (Fig.~\ref{fg4}). The line is weak and present only in the
8 exposures of the frame 580l. 
Because of the similar sensitivity to $\alpha$ changes, its position should coincide with the 
\ion{Fe}{i} $\lambda2484$ \AA\ and \ion{Fe}{i} $\lambda2523$ \AA\ lines,
but the cross-correlation showed that the \ion{Fe}{i} $\lambda3720$ \AA\ line was offset by 
$-0.3$ \kms. 
On the other hand, the profiles of \ion{Fe}{i} $\lambda\lambda2484,2523, 2967$, and 3021 (corrected) 
were coherent which means that \ion{Fe}{i} $\lambda3720$ was affected by some miscalibration.
Thus, the final line profile of \ion{Fe}{i} $\lambda3720$ \AA\ was shifted by 0.3 \kms.

\ion{Fe}{ii} $\lambda2586$ \AA\ (Fig~\ref{fg5}, right panel). 
This quite strong line is blended with Ly-$\alpha$ forest absorptions, but can be accurately
deconvolved. The noise was calculated as the difference between the observed and the fitted spectrum
and added to the deconvolved \ion{Fe}{ii} $\lambda2586$ \AA\ profile shown in Fig.~\ref{fg5}, left panel.

\ion{Ca}{i} $\lambda4227$ \AA\ (Fig.~\ref{fg6}). 
This weak line is located at $\lambda = 6138$ \AA. Close to its position, there are 
a weak \ion{Fe}{ii} $\lambda2374$ \AA\ 
(at $\lambda = 6142$ \AA), and a strong \ion{Fe}{ii} $\lambda2382$ \AA\ (at $\lambda = 6163$ \AA) lines, both
from the \zabs = 1.5864 system. 
The lines are present in the 14 exposures: 8 from the frame 580u, 3 from 700l, all 11 with a similar signal,
and 3 from the frame 760l with high but different in all 3 exposures signals.
The velocity shifts of the profiles from the frame 760l
relative to the co-added 11 profiles from the frames 580u and 720l
were calculated for the high-contrast \ion{Fe}{ii} $\lambda2382$ \AA\ line. 
Then the same shifts were applied to the \ion{Fe}{ii} $\lambda2374$ \AA\ exposures from the setting 760l.
The final profiles of \ion{Fe}{ii} $\lambda\lambda2382, 2374$ \AA\ were perfectly
aligned with the \ion{Fe}{ii} $\lambda\lambda2600, 2586, 1608$ \AA\ lines 
and with other low-ionization lines from the  \zabs\ = 1.5864 system. 
The same procedure was applied to 
the \ion{Ca}{i} line: profiles from the frame 760l were shifted in accord with 
the \ion{Fe}{ii} $\lambda2382$ \AA\ exposures and then
co-added with unchanged profiles from the frames 580u and 700l. 

\ion{Mg}{i} $\lambda2852$ \AA\ (Fig.~\ref{fg6}). The line is weak and located at 4142 \AA. Close to its position 
at 4150 \AA, there is a high-contrast line of
\ion{O}{i} $\lambda1032$ \AA\ from the \zabs = 2.1871 system (Sect.~\ref{sect-4-4}). 
The \ion{Mg}{i} profiles from the high-signal exposures (settings 437) were 
corrected  following the shifts of this \ion{O}{i} line.

\ion{Mg}{ii} $\lambda\lambda2796, 2803$ \AA\ (Fig.~\ref{fg6}). 
Both lines are strong and present in all 14 exposures. However, the
profiles of \ion{Mg}{ii} $\lambda2803$ \AA\ in 4 exposures (2 of the setting 390, 
1 of the setting 437, and 1 of the setting 420) 
are severely distorted and not included in the final spectrum. This explains different
S/N ratios for the \ion{Mg}{ii} $\lambda2796$ \AA\ and \ion{Mg}{ii} $\lambda2803$ \AA\ lines 
shown in Fig.~\ref{fg6}.

As a result, we have for the analysis the following set of the absorption lines:
\ion{Mg}{i} $\lambda2852$, \ion{Mg}{ii} $\lambda\lambda2796, 2803$, 
\ion{Fe}{i} $\lambda\lambda2484, 2523, 2967, 3021, 3720$,
\ion{Fe}{ii} $\lambda\lambda2600, 2382, 2586, 2344, 2374$,
\ion{Ca}{i} $\lambda$4227, and \ion{Ca}{ii} $\lambda\lambda3934, 3969$ \AA. 
The line profiles of each ion were analyzed independently. 
Trial calculations have shown that
a single-component Gaussian profile is sufficient to fit properly all lines.
However, the value of $\chi^2 \la 1$ in the fitting of the \ion{Mg}{ii} doublet (highest
S/N ratio) could be achieved only with a covering factor ${\cal C} < 1$, 
i.e., assuming an incomplete coverage
of the background light source: alternative models with two components and
${\cal C} = 1$ yield very
small $b$-parameter ($\approx 0.5$ km/s) which is 
inconsistent  with the line widths of
other species from this system. Spectra of other ions have lower S/N ratios 
and their fits with ${\cal C} < 1$
and with ${\cal C} = 1$ are statistically indistinguishable, although $\chi^2$ 
in trials with ${\cal C} < 1$
was always lower (e.g., 0.7 vs. 0.9). Thus, the final calculations were performed 
with ${\cal C} < 1$ for all ions.

The covering factor ${\cal C}$ is easily determined from a slightly modified 
model of the line profile, $I_\lambda = {\cal C}\exp(-\tau_\lambda) + (1 - {\cal C})$,
under the assumption that ${\cal C}$ is the same for the whole absorbing cloud
(i.e., it does not depend on $\lambda$ or, equivalently, on the radial velocity $V$). 
Here,  $I$ is the observed normalized intensity within the
absorption line profile and $\tau_\lambda$ is the optical depth. 
The covering factor can differ from ion to ion since, in general, different ions trace different gas. 
If at least two lines of the same ion are
available, then the covering factor is estimated safely. If only one line is present, then
the covering factor cannot be determined (degenerate with column density) and has to be fixed to some value 
(see note to Table~\ref{tbl-4}).
In fact, the incomplete covering is quite common among the so-called associated (i.e. located close
to the background quasar) absorbers and not such a rare
finding even among intergalactic  absorbers (see examples in Paper~I).   
The cause of the light leakage is not clear, but, probably, it is related to the 
gravitational lensing of the quasar beam by some intervening galaxy/cluster.
In any case we can conclude that the absorber showing the incomplete covering of the
background source is to be very small, probably even of a sub-parsec linear size. 
Note that the incomplete covering in the \zabs\ = 0.45207 system was recently mentioned also 
by Jones \etal\ (2010).

The observed and synthetic profiles from the \zabs\ = 0.45207 system
are shown in Figs.~\ref{fg4}-\ref{fg6}, 
and the corresponding fitting parameters are listed in Table~\ref{tbl-4}.

\begin{figure}[t]
\vspace{-3.0cm}
\hspace{-4.0cm}\psfig{figure=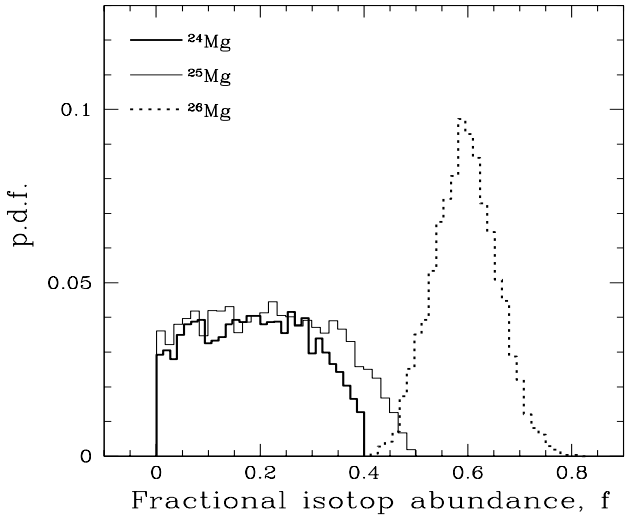,height=14cm,width=16cm}
\vspace{-3.5cm}
\caption[]{Empirical probability distribution functions (p.d.f.) 
for Mg isotopic fractional abundances, $f$, estimated by
a Monte Carlo procedure (see Sect.~\ref{sect-4-1}, for detail) }
\label{fg8}
\end{figure}

The radial velocity of the system center calculated as the weighted mean value of the \ion{Fe}{ii}, \ion{Fe}{i}, 
\ion{Ca}{ii}, and \ion{Ca}{i} line centers is 
$V_{\rm c} = -1.77\pm0.03$ \kms. Then for the velocity offsets of the Mg lines we
obtain the following values: 
$\Delta V_{\scriptscriptstyle \rm Mg\,{\sc II}}
\equiv V_{\scriptscriptstyle \rm Mg\,{\sc II}} - V_{\rm c} = -0.44\pm0.05$ \kms,     
$\Delta V_{\scriptscriptstyle \rm Mg\,{\sc I}}
\equiv V_{\scriptscriptstyle \rm Mg\,{\sc I}} - V_{\rm c} = -0.17\pm0.17$ \kms,     
i.e., negative velocity shifts are detected for both the \ion{Mg}{ii} lines and the \ion{Mg}{i} line 
although for the latter it is only at the $1\sigma$ level because of a large uncertainty of the \ion{Mg}{i} line
position (weak single line). 
However, before ascribing this shift to the enhanced content of the heavy Mg isotopes 
in the absorbing gas, the
arguments concerning possible calibration errors and kinematic shifts (the so-called `Doppler noise') 
should be considered.

The centers of both \ion{Fe}{ii} and \ion{Fe}{i} lines are determined from 5 transitions, 
and therefore they provide a robust mean radial velocity.
The calibration of \ion{Ca}{ii} and \ion{Ca}{i} lines 
is verified by 
the \ion{Fe}{ii} $\lambda\lambda2382, 2374$ \AA\ lines from the \zabs\ = 1.5864 system
and their centers coincide with those of the \ion{Fe}{ii} and \ion{Fe}{i} lines within 
the $1\sigma$ uncertainty interval.

Both \ion{Mg}{ii} lines fall in the
spectral region with the highest S/N ratio. 
They are narrow and strong~-- this explains a small error (0.04 \kms) of the line center determination.
However, the systematic error due to miscalibration can easily reach the tenfold value
as was shown above for, e.g., the \ion{Fe}{i} $\lambda3021$ \AA\ and 
\ion{Fe}{i} $\lambda3720$ \AA\ lines. 
Can the positions of the \ion{Mg}{ii} lines be also
affected by some additional errors~? The answer is very probable negative. 
The main factor here is that the
position of \ion{Mg}{ii} $\lambda2803$ coincides with a strong and unsaturated line 
\ion{Th}{i} $\lambda4070$ \AA\ from the ThAr spectrum
which is used as a
reference for the wavelength calibration. As a result of this coincidence,
the \ion{Mg}{ii} $\lambda2803$ profiles from different exposures are almost aligned 
(shifts $< 0.05$ \kms). Individual profiles
of the \ion{Mg}{ii} $\lambda2796$ reveal shifts up to $\pm250$ \ms\ (see, e.g., Fig.~\ref{fg2}), 
but the final \ion{Mg}{ii} $\lambda2796$ profile
prepared by the standard procedure (i.e., similar-signal exposures co-added to
form a preliminary profile, high-signal exposures aligned with this profile
and then all exposures co-added again) is fully coherent with \ion{Mg}{ii} $\lambda2803$.
In general, all lines in the present study which were found to be shifted due to calibration errors
(e.g., \ion{Fe}{i} $\lambda3021$, this Section, \ion{Fe}{ii} $\lambda1608$ at \zabs\ = 2.1871, 
Sect.~\ref{sect-4-4}) 
fall in the gaps between ThAr reference lines, whereas lines with good calibration 
(e.g., \ion{Si}{ii} $\lambda1526$ and \ion{Fe}{ii} $\lambda2344$ from the
\zabs\ = 2.1871 system, Sect.~\ref{sect-4-4}) coincide with some of the ThAr lines. 
The conclusion that the vicinity to some of the ThAr
reference lines ensures a stable wavelength calibration is further supported by the comparison of the line
positions in the archive \object{HE0001--2340} spectrum and in the present one: the `coinciding' lines are coherent
in both spectra whereas centers of lines from the gaps can differ by $\sim 0.5$ \kms. 
Thus, the negative shift of the \ion{Mg}{ii}
lines in the system under consideration cannot be caused by calibration errors. The same statement is
valid for the \ion{Mg}{i} $\lambda2852$ line as well, since its position is also close to a ThAr reference.

\begin{figure*}[t]
\vspace{0.0cm}
\hspace{0.0cm}\psfig{figure=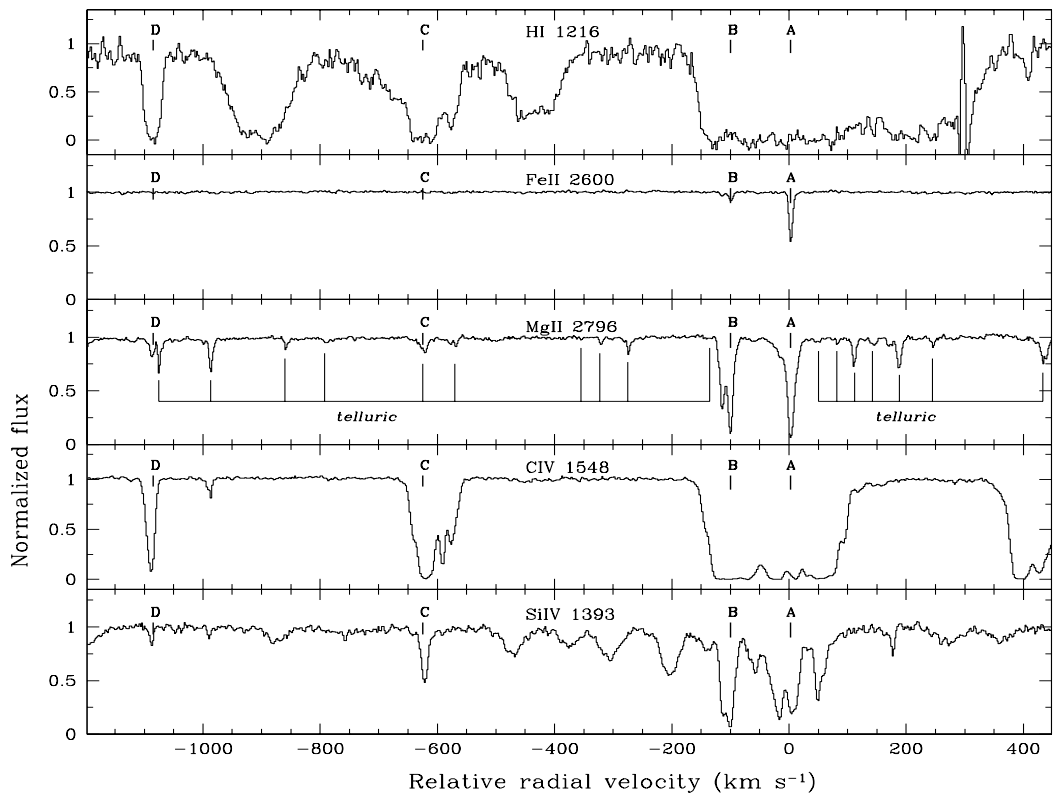,height=14cm,width=16cm}
\vspace{-3.0cm}
\caption[]{
The overview of the H\,{\sc I} Ly-$\alpha$, Fe\,{\sc II} $\lambda2600$ \AA, Mg\,{\sc II} $\lambda2796$ \AA,
C\,{\sc IV} $\lambda1548$ \AA, and Si\,{\sc IV} $\lambda1393$ \AA\  
profiles from the \zabs = 1.5865 absorption complex 
towards the quasar \object{HE0001--2340}.
The  complex consists of four sub-systems ($A, B, C, D$) spread over 1100 \kms.
The zero radial velocity is fixed at $z_A = 1.5864$.
}
\label{fg9}
\end{figure*}

Neither can this shift be attributed to the gas flows inside the absorber. 
The estimated Doppler $b$-parameters (Table~\ref{tbl-4}) show that 
the line broadening is not thermal and, hence, some  turbulent
component is present. 
But the line centers of \ion{Fe}{ii} and \ion{Fe}{i} (as well as \ion{Ca}{ii} and \ion{Ca}{i}) 
coincide within the $1\sigma$ uncertainty interval~-- contrary to the centers of the \ion{Mg}{ii} and
\ion{Mg}{i} lines. 
However, at values of the ionization parameter
$\log U \la -4$, which is the range most probable
for the system under study (D'Odorico 2007; Jones \etal\ 2010), 
the ionization curves of the \ion{Mg}{ii} and \ion{Fe}{ii} ions
are almost parallel whereas the ionization
curves of \ion{Fe}{i} and \ion{Fe}{ii} diverge. Thus, in case of any density/velocity gradients inside the
absorbers one would expect rather a shift between the centers of the \ion{Fe}{i} and \ion{Fe}{ii} lines but
not between \ion{Fe}{ii} and \ion{Mg}{ii}.

It remains to assume that the measured negative shift of the \ion{Mg}{ii} and \ion{Mg}{i} lines 
relative to other lines in the \zabs\ = 0.45207
system is indeed due to enhanced content of heavy Mg isotopes in the absorbing gas.
Figure~\ref{fg7} shows the velocity shift $\Delta V$ as a function of the ratio  
$\xi =\ $(\mgb + \mgc)/\mga 
for the \ion{Mg}{i} $\lambda2852$ \AA\ (upper panel) and
the \ion{Mg}{ii} $\lambda2796$ \AA\ (lower panel) lines.
The curves are calculated for different ratios
of the heavy isotopes, $r =\ $\mgc/\mgb, which are depicted over each curve.
The laboratory wavelengths 
for separate isotopes are taken from Salumbides \etal\ (2006) and
from Batteiger \etal\ (2009) for, respectively, \ion{Mg}{i} and \ion{Mg}{ii}.
$\Delta V$ = 0  corresponds to the solar value of the Mg isotope
ratios, $\xi_\odot = 0.2660$ and $r_\odot \approx 1$.  
The horizontal dashed lines mark the measured mean values of 
$\Delta V_{\scriptscriptstyle \rm Mg\,{\sc I}} = -170$ \ms\
and $\Delta V_{\scriptscriptstyle \rm Mg\,{\sc II}} = -440$ \ms, 
and the dotted lines restrict the $\pm1\sigma$
uncertainty interval for \ion{Mg}{ii} (for \ion{Mg}{i}, it runs from 
$\Delta V_{\scriptscriptstyle \rm Mg\,{\sc I}} = 0$ to $-340$ \ms).
From the lower panel of Fig.~\ref{fg7} we see immediately that the 
measured shift of the \ion{Mg}{ii} lines can be
produced by the isotope mixture with $\xi > 1.5$ and $r > 1$, i.e., with an overabundance of
\mgc\ relative to other two isotopes.

\begin{figure*}[t]
\vspace{0.0cm}
\hspace{0.0cm}\psfig{figure=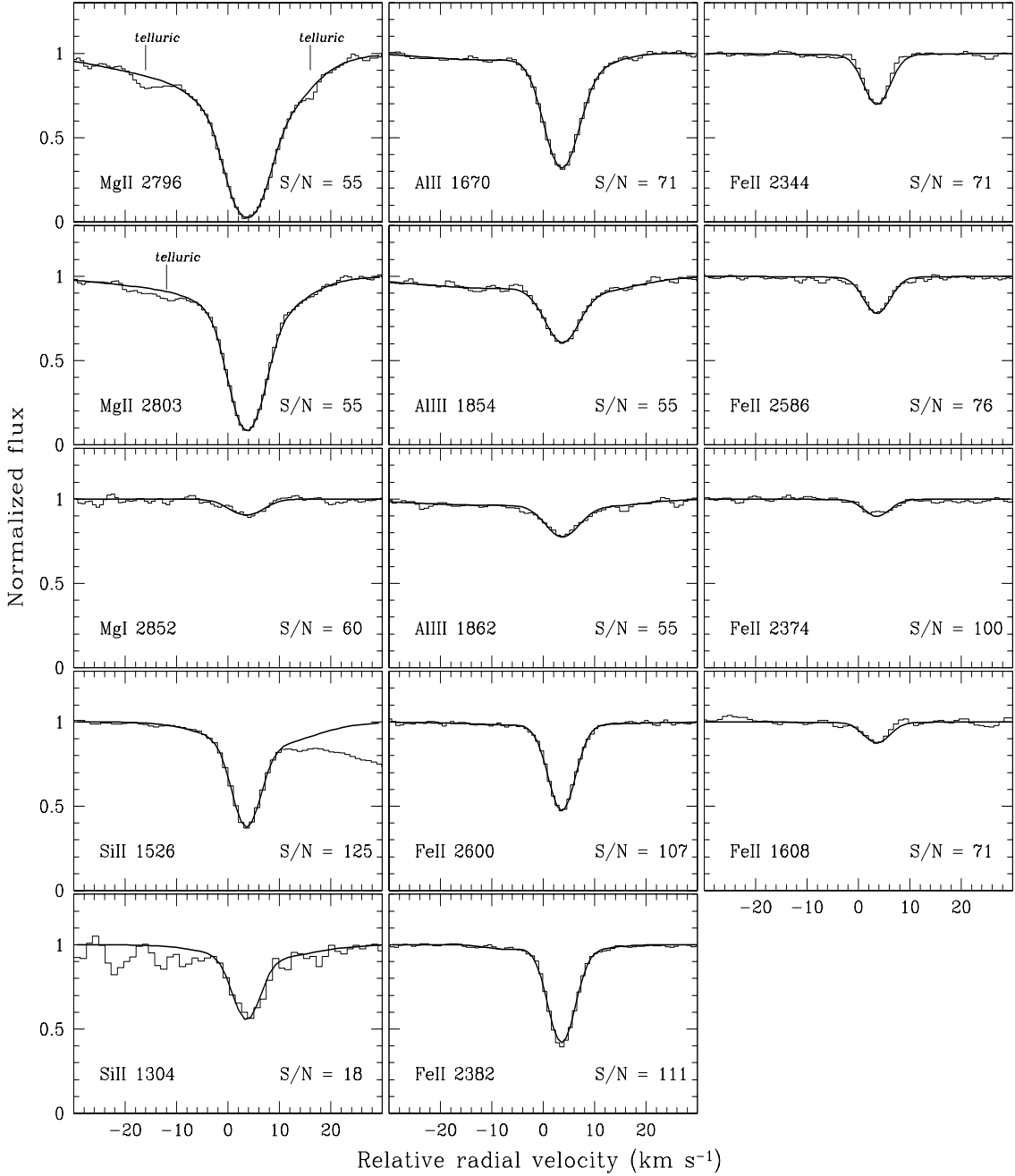,height=18cm,width=16cm}
\vspace{0.0cm}
\caption[]{
Metal absorption lines from
the \zabs = 1.5864 system towards the quasar \object{HE0001--2340}
(solid-line histograms).
Synthetic profiles are plotted by the smooth curves.
The signal-to-noise ratio, S/N, per pixel in the co-added spectrum is indicated in each panel.
The zero radial velocity is fixed at $z = 1.5864$.
}
\label{fg10}
\end{figure*}

Using the values of $\Delta V_{\scriptscriptstyle \rm Mg\,{\sc I}}$ and 
$\Delta V_{\scriptscriptstyle \rm Mg\,{\sc II}}$ and their errors, we can estimate
the most probable fractional abundances $f_{24}, f_{25}$, and $f_{26}$
of Mg isotopes assuming that the measured
velocity shifts are random values normally distributed with the means $-170$ \ms\
and $-440$ \ms\ and the dispersions of 170 \ms\ and 50 \ms, respectively.
Fig.~\ref{fg8} shows the empirical probability distribution functions
for each isotope obtained by Monte Carlo modeling when 
$\Delta V_{\scriptscriptstyle \rm Mg\,{\sc I}}$ and
$\Delta V_{\scriptscriptstyle \rm Mg\,{\sc II}}$
were chosen at random from the corresponding normal distributions
and $f_{24}, f_{25}$, and $ f_{26}$ were calculated under the conditions  
$f_{24} + f_{25} + f_{26} = 1$, each $f_k > 0$,
$\xi = (f_{25} + f_{26})/f_{24} > 1.5$, and $r = f_{26}/f_{25} > 1$.
It is seen that the abundances of the first two isotopes \mga\ and \mgb\ are almost
rectangularly distributed, whereas
the fractional abundance of \mgc\ is distributed normally.
The calculated statistics are $f_{24} = 0.19\pm0.11$, $f_{25} = 0.22\pm0.13$, and
$f_{26} = 0.59\pm0.06$.

\begin{table}[t!]
\centering
\caption{Fitting parameters of spectral lines identified in the \zabs\ = 1.5864 absorber.
For each ion, the upper row lists parameters of the main component whereas 
parameters of auxiliary components are listed below.
Indicated are $1\sigma$ statistical errors. } 
\label{tbl-5}
\begin{tabular}{l r@{.}l r@{.}l r@{.}l }
\hline
\hline
\noalign{\smallskip}
\multicolumn{1}{c}{Ion} & \multicolumn{2}{c}{$V$, \kms} & 
\multicolumn{2}{c}{$b$, \kms} & \multicolumn{2}{c}{$N$, \cm} \\
\noalign{\smallskip}
\hline
\noalign{\smallskip}
Fe\,{\sc ii} & 3&$56\pm0.05$ & 1&$92\pm0.04$ & (3&$48\pm0.03$)E12 \\[-1pt]
           &$-6$&$75\pm1.26$ & 4&$5\pm1.0$   & (1&$28\pm0.12$)E11 \\[-1pt]
           & 9&$59\pm0.30$  & 5&$2\pm1.6$ & (1&$34\pm0.12$)E11 \\[2pt]
Si\,{\sc ii} & 3&$52\pm0.05$ & 2&$09\pm0.10$ & (1&$03\pm0.01$)E13 \\[-1pt]
           &$-5$&$2\pm5.5$ & 19&$0\pm1.5$   & (8&$69\pm1.16$)E11 \\[-1pt]
 & 5&$7\pm0.3$  & 8&$8\pm0.1$ & (6&$41\pm0.15$)E12 \\[2pt]
Al\,{\sc ii} & 3&$55\pm0.15$ & 3&$05\pm0.30$ & (1&$15\pm0.05$)E12 \\[-1pt]
           &$-12$&$0\pm2.1$ & 14&$1\pm1.2$   & (1&$3\pm0.1$)E11 \\[-1pt]
 & 7&$4\pm1.1$ & 7&$4\pm1.0$ & (1&$78\pm0.18$)E11 \\[2pt]
Al\,{\sc iii} & 3&$65\pm0.36$ & 3&$24\pm0.30$ & (1&$19\pm0.04$)E12 \\[-1pt]
 & 10&$41\pm1.51$ & 11&$57\pm2.02$ & (5&$28\pm0.10$)E11 \\[-1pt]
 & $-11$&$08\pm5.32$ & 22&$48\pm5.20$ & (1&$02\pm0.06$)E12 \\[2pt]

Mg\,{\sc ii} & 3&$71\pm0.04$ &  2&$79\pm0.02$ & (9&$10\pm0.10$)E12 \\[-1pt]
 & $-5$&$81\pm0.53$ & 21&$13\pm0.66$ & (1&$43\pm0.04$)E12 \\[-1pt]
 &  $5$&$23\pm0.10$ & 9&$65\pm0.11$ & (2&$25\pm0.03$)E12 \\[2pt]
Mg\,{\sc i} & 3&$58\pm0.45$ &  3&$74\pm0.82$ & (6&$1\pm0.6$)E10 \\
\noalign{\smallskip}
\hline
\noalign{\smallskip}
\end{tabular}
\end{table}

In order to understand what causes the enhanced abundances of heavy Mg isotope  in 
the \zabs\ = 0.45207 system the origin of this absorber should be clarified. 
Consider now the column densities of the ions presented in Table~\ref{tbl-4}. 
Although accurate
ionization corrections are not known, we can unambiguously conclude that
the relative abundance of iron to magnesium, (Fe/Mg) $\ga 5$, 
is by almost an order of magnitude higher than its solar value, (Fe/Mg)$_\odot = 0.83$ since the
ionization corrections for \ion{Fe}{ii} are always higher than that for \ion{Mg}{ii} irrespectively
of the shape of the ionizing radiation and the value of the ionization parameter $U$.
The ionization curves of magnesium and calcium at $\log U \la -4$ closely trace each other, and 
from the comparison of the column densities of
\ion{Mg}{i} and \ion{Mg}{ii} with those of \ion{Ca}{i} and \ion{Ca}{ii} we obtain
that the relative content of calcium to magnesium, (Ca/Mg) $\approx 0.4$, 
considerably exceeds its solar value, (Ca/Mg)$_\odot = 0.06$.
As already mentioned above, in the system under study there is probably the 
\ion{Si}{i} $\lambda2515$ \AA\ line, 
but its identification is uncertain since it falls into the Ly-$\alpha$ forest. 
If the absorption at the expected position of the \ion{Si}{i} line is indeed due to this transition, 
then $N$(\ion{Si}{i}) = $(7.0\pm0.7)\times10^{11}$ \cm\ (D'Odorico 2007).
The ionization correction for \ion{Si}{i} at $\log U \la -4$ is very close to that for \ion{Mg}{i},
and comparing $N$(\ion{Si}{i}) with $N$(\ion{Mg}{i})
we obtain again a significant relative overabundance of 
silicon to magnesium: (Si/Mg) = 3.7 vs. (Si/Mg)$_\odot = 0.98$.
Such an abundance pattern~-- a high iron content, overabundances of Si and Ca relative to Mg~--
is usually observed in remnants of type Ia supernovae
(e.g., Stehle \etal\ 2005; Mazzali \etal\ 2008; Tanaka \etal\ 2010). 
The \ion{Mn}{ii} triplet, although indistinguishable in our data, is clearly seen in the
low-resolution spectrum of \object{HE0001--2340} where
$N$(\ion{Mn}{ii}) $=(3.6\pm0.4)\times10^{11}$ \cm\ was determined by D'Odorico (2007). 
It is not clear how to translate this column density into the Mn abundance,
but taking into account the trace solar content of Mn, (Mn/Fe)$_\odot = 0.01$, 
a very high relative abundance and even an overabundance of Mn relative to Fe seems to be quite probable.
And this supposes a high metallicity progenitor, since,
in general, the abundance of Mn in the supernova Ia remnants is thought to be correlated with the metallicity 
(expressed as the initial content of $^{22}$Ne) of
the exploding white dwarf (McWilliam \etal\ 2003;  Badenes \etal\ 2008).
Model calculations of nucleosynthetic yields from SN Ia 
do not predict enhanced content of heavy Mg isotopes for the
whole SN Ia remnant, but very high concentration of \mgc\ 
is predicted for the outer layer of the exploding white dwarf
where it is synthesized via alpha-capture on $^{22}$Ne 
(Seitenzahl \etal\ 2010). 
The SN remnants are known
to be heterogeneous objects displaying complex variations of chemical
composition with position and velocity (e.g., Hughes \etal\ 2000). 
In fact, the \zabs\ = 0.45207 system also consists of two separated components
(Fig.~\ref{fg3}) with obviously different physical parameters. 
Taking into account a tiny linear size (manifested through the incomplete coverage of
the light source) of the analyzed clump at $V = 0$ \kms\ 
we can suggest that it is a fragment of a SN Ia remnant.

As already mentioned above, different sensitivity of different \ion{Fe}{i}, \ion{Fe}{ii}, 
and \ion{Ca}{ii} transitions to changes in the fine-structure constant $\alpha$
makes it possible to estimate $\Delta \alpha/\alpha$ from the velocity shifts of 
the corresponding line centers.
Unfortunately, the measured errors of the line centers in the present system allows us to put constraints on 
$\Delta \alpha/\alpha$  only at the level of $10^{-5}$,
with the most accurate estimate provided by \ion{Fe}{i} lines:
$\Delta\alpha/\alpha = (0.6\pm1.0)\times10^{-5}$.

We note that Chand \etal\ (2004) and Murphy \etal\ (2008) estimated $\Delta \alpha/\alpha$ in this system
using the \ion{Mg}{ii} $\lambda\lambda2796, 2803$ \AA\ and 
\ion{Mg}{i} $\lambda2852$ \AA\ transitions as the reference. 
A negative $\Delta \alpha/\alpha$ value
reported by Murphy \etal\ is obviously a consequence of the unaccounted Mg isotope shift.

\begin{figure}[t]
\vspace{0.0cm}
\hspace{0.0cm}\psfig{figure=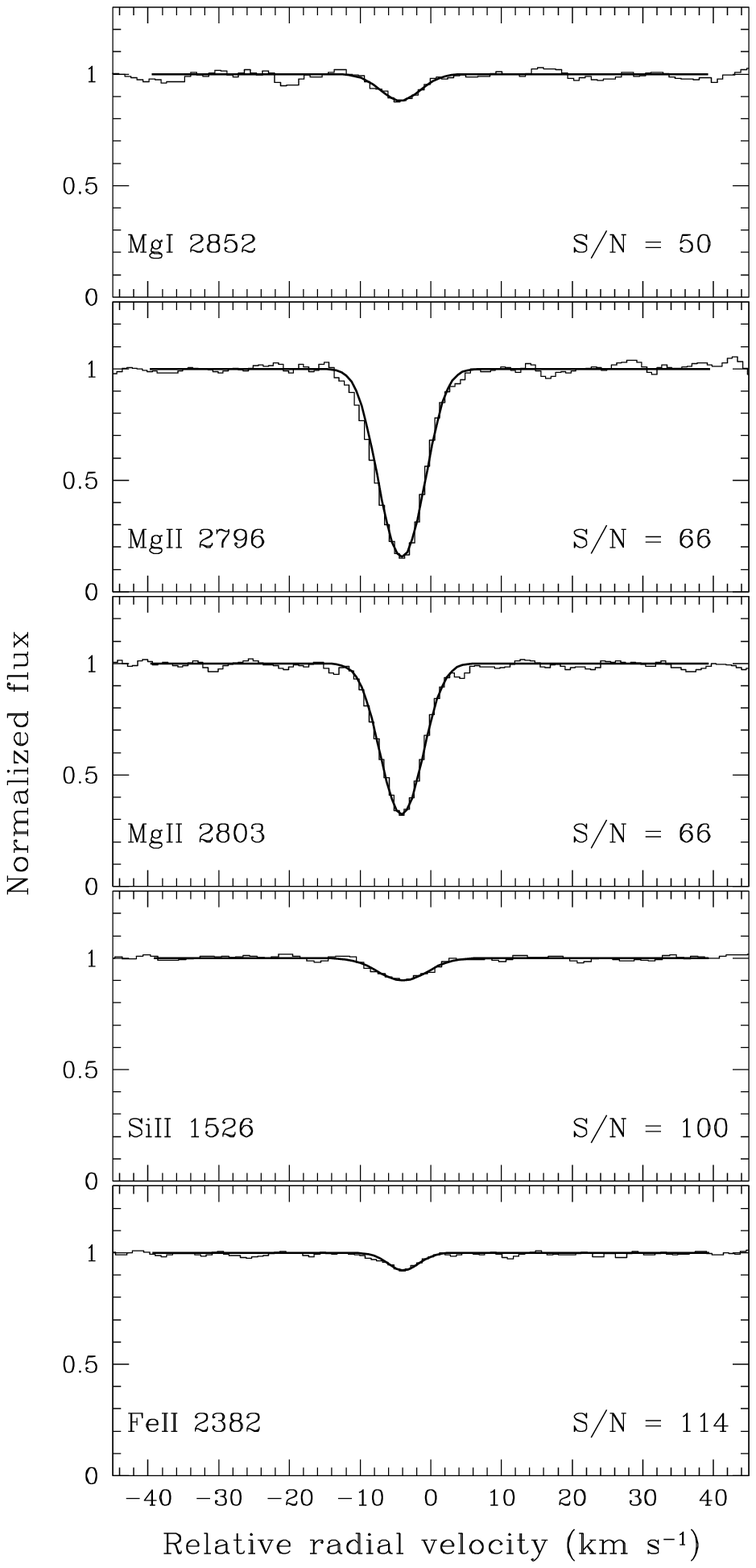,height=18cm,width=16cm}
\vspace{-0.5cm}
\caption[]{
Metal absorption lines from
the \zabs = 1.6515 system towards the quasar \object{HE0001--2340}
(solid-line histograms).
Synthetic profiles are plotted by the smooth curves.
The signal-to-noise ratio, S/N, per pixel in the co-added spectrum is indicated in each panel.
The zero radial velocity is fixed at $z = 1.6515$.
}
\label{fg11}
\end{figure}

\begin{table}[t!]
\centering
\caption{Fitting parameters of spectral lines identified in the \zabs\ = 1.6515 absorber.
Indicated are $1\sigma$ statistical errors.
} 
\label{tbl-6}
\begin{tabular}{l r@{.}l r@{.}l r@{.}l }
\hline
\hline
\noalign{\smallskip}
\multicolumn{1}{c}{Ion} & \multicolumn{2}{c}{$v$, \kms} & \multicolumn{2}{c}{$b$, \kms} & \multicolumn{2}{c}{$N$, \cm} \\
\noalign{\smallskip}
\hline
\noalign{\smallskip}
Si\,{\sc ii} & -4&$00\pm0.30$ & 4&$0\pm1.4$ & (1&$75\pm0.14$)E12 \\[2pt]
Fe\,{\sc ii} & -4&$10\pm0.30$ &  2&$2\pm1.0$ & (2&$35\pm0.50$)E12 \\[2pt]
Mg\,{\sc ii} & -4&$13\pm0.03$ &  2&$53\pm0.03$ & (4&$00\pm0.04$)E12 \\[2pt]
Mg\,{\sc i} & -4&$33\pm0.39$ &  2&$7\pm1.5$ & (6&$46\pm1.42$)E10 \\[2pt]
\noalign{\smallskip}
\hline
\end{tabular}
\end{table}

\subsection{Absorber at \zabs\ = 1.5864}
\label{sect-4-2}

This system represents a large absorbing complex extending over  $\sim1000$ \kms\ (Fig.~\ref{fg9}). 
The redshift $z = 1.5864$ 
corresponds to the sub-system $A$ with the strongest low-ionization lines (\ion{Fe}{ii}, \ion{Si}{ii}, \ion{Mg}{ii},
\ion{Al}{ii}). 
The sub-system $D$ at $V = -1200$ \kms\ 
was described in Paper~I as a high-metallicity absorber 
enriched by AGB-stars.
The sub-systems $A$ and $B$ probably originate in low-ionization clumps (lines
of \ion{F}{ii}, \ion{Mg}{ii}, \ion{Si}{ii}, \ion{C}{ii}, \ion{Al}{ii}) 
embedded in gas with a higher ionization which is seen in lines of \ion{Si}{iv} and \ion{C}{iv}.
Note a remarkable kinematic similarity between \ion{Mg}{ii} and \ion{Fe}{ii} 
absorptions in these sub-systems
and \ion{Mg}{ii} and \ion{Fe}{ii} absorptions in the previously described system at \zabs\ = 0.45207
(see Fig.~\ref{fg3}).

Here we describe the sub-system $A$ in the context of measuring
the Mg isotope abundance ratio. Unfortunately, in the sub-system $B$ 
the \ion{Mg}{ii} $\lambda\lambda2796, 2803$ lines are contaminated with
telluric absorptions
which prevent to measure accurately their centers. 
The observed line profiles shown by histograms in Fig.~\ref{fg10} 
were prepared as described in Sect.~\ref{sect-3}. 
Some particular details are given below.

\ion{Fe}{ii} $\lambda2586$ \AA. Three profiles from the frame 580u 
are severely corrupted and were excluded from
the final co-added spectrum. This explains a lower S/N at the position of this line
as compared to the neighboring \ion{Fe}{ii} $\lambda2600$ \AA\ line.

\ion{Fe}{ii} $\lambda2600$ \AA. The cross-correlation 
of the \ion{Fe}{ii} $\lambda2600$ \AA\ and \ion{Fe}{ii} $\lambda2382$ \AA\ profiles 
revealed a shift of 0.15 \kms\ between them. Since trial fittings of 
\ion{Fe}{ii} lines have shown the consistency of \ion{Fe}{ii} $\lambda2382$ \AA\ with 
\ion{Fe}{ii} $\lambda\lambda2344, 2586, 2374$, and 1608 \AA,
the final \ion{Fe}{ii} $\lambda2600$ \AA\
line was aligned with \ion{Fe}{ii} $\lambda2382$ \AA\ (i.e., shifted by 0.15 \kms).  

\ion{Fe}{ii} $\lambda1608$ \AA. 
The line is very weak and barely visible in the individual exposures. 
In the spectrum of \object{HE0001--2340}, it lies at 
4160 \AA~-- close to a strong \ion{O}{i} $\lambda1302$ \AA\ line from the \zabs\ = 2.1871 system 
at 4150 \AA\ (Sect.~\ref{sect-4-4}). 
The position of this \ion{O}{i} line is
verified by comparison with the \ion{Si}{ii} $\lambda1260$ \AA\ and 
\ion{Si}{ii} $\lambda1526$ \AA\ lines from  the \zabs\ = 2.1871 system. 
Thus, the \ion{Fe}{ii} $\lambda1608$ \AA\ profiles
from 3 exposures of the high-signal setting 437 were shifted 
in accord with the shifts of the \ion{O}{i} $\lambda1302$ \AA\ line.

\ion{Al}{ii} $\lambda1670$ \AA. See Sect.~\ref{sect-4-1} where \ion{Fe}{i} $\lambda2967$ \AA\ line is described.

\ion{Mg}{ii} $\lambda\lambda2796, 2803$ \AA. Both lines are very strong and present in  
3 exposures of the frame 760l and 3 exposures of the frame 700u,
all having different S/N ratios. The exposure with the highest S/N (\#8 from 700u, Table~\ref{tbl-1})
was taken as a reference and all other exposures were aligned with it before co-adding.  

\ion{Mg}{i} $\lambda2852$ \AA. The line is very weak and its profiles in the
individual exposures (3 from  760l and 3 from  700u) 
are severely distorted by noise. All available exposures were simply
co-added without any preliminary alignments. This means that the final profile is mostly affected
by the strongest exposure \#8 from  700u.

\ion{C}{ii} $\lambda1334$ \AA. The line is almost saturated, has a low S/N and is partly blended with
Ly-$\alpha$ forest absorptions. The line center cannot be estimated with accuracy better
than 1 \kms\ and we did not include the \ion{C}{ii} $\lambda1334$ \AA\ line in the final
analysis.

The lines of each ion (\ion{Fe}{ii}, \ion{Si}{ii}, \ion{Al}{ii}, \ion{Al}{iii}, \ion{Mg}{ii}, \ion{Mg}{i})  
were fitted independently using Gaussian components. In each case
the number of components was determined as a minimum required to fit the profile with $\chi^2 \la 1$.
The fitting parameters are listed in Table~\ref{tbl-5}, the synthetic profiles are shown in Fig.~\ref{fg10} by  
the solid lines. 
The weighted value for the system center calculated from
the \ion{Fe}{ii}, \ion{Si}{ii}, and \ion{Al}{ii} lines 
is $V_c = 3.54 \pm 0.03$ \kms. This gives the offset for the \ion{Mg}{ii} line of  
$\Delta V_{\scriptscriptstyle \rm Mg\,{\sc II}} 
\equiv 
V_{\scriptscriptstyle \rm Mg\,{\sc II}} - V_c = 0.17 \pm 0.05$ \kms\ and 
$\Delta V_{\scriptscriptstyle \rm Mg\,{\sc I}} = 0.04 \pm 0.45$ \kms\ for \ion{Mg}{i}.

\begin{figure}[t]
\vspace{-1.0cm}
\hspace{-2.8cm}\psfig{figure=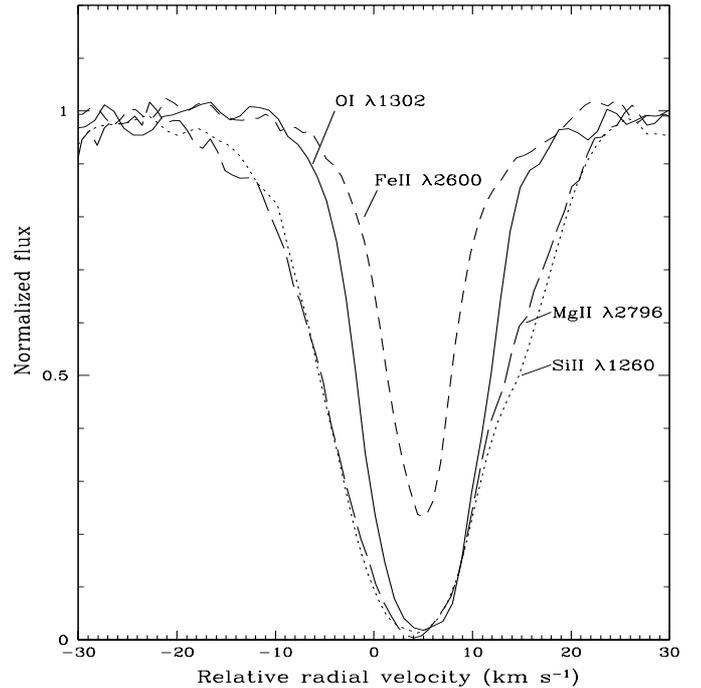,height=12cm,width=15cm}
\vspace{-1.0cm}
\caption[]{
Absorption lines from the \zabs\ = 2.1871 system overlapped to demonstrate different profile shapes.
The \ion{Fe}{ii} $\lambda2600$ \AA\ line is shifted by 0.35 \kms, and the
\ion{Mg}{ii} $\lambda2796$ \AA\ line by 1.08 \kms\ with respect to their
original positions in the spectrum (see Sect.~\ref{sect-4-4}, for detail). 
The zero radial velocity is fixed at $z = 2.1871$.
}
\label{fg12}
\end{figure}

As mentioned above, the profiles of the 
\ion{Mg}{ii} lines from individual exposures were aligned with the exposure 
\#8 from the frame 700u having the strongest signal. The offset revealed is the sum of two effects: a
shift of the reference exposure, and a shift 
of the \ion{Mg}{ii} lines due to enhanced content of heavy isotopes.
The value of the exposure shift can be estimated from the following consideration.
The \ion{Mg}{ii} lines are located in the range  
$\Delta \lambda = 7230 - 7252$ \AA\ . 
The frame 700u starts at 7050 \AA\ (Table~\ref{tbl-2}), 
and up to the positions of the \ion{Mg}{ii} lines there are 
no absorption lines in the spectrum which can be used to verify the wavelength calibration. 
However, the frame 760l (another frame whose exposures contribute to the \ion{Mg}{ii} profiles) 
starts at shorter wavelengths (covering
$\Delta \lambda =  5694 - 7532$ \AA), and up to 6809 \AA\ it overlaps with the wavelength interval 
covered by the frame 580u.
In the range covered by both the 760l and 580u frames there are several absorption lines 
from three different systems (\ion{Ca}{ii} $\lambda\lambda3934, 3969$ \AA, \ion{Ca}{i} $\lambda4227$ \AA\ at
\zabs\ = 0.45207; \ion{Fe}{ii} $\lambda\lambda2344, 2374, 2382, 2586, 2600$ \AA\ from the present system;
and \ion{Fe}{ii} $\lambda2382$ \AA\ at  \zabs\ = 1.6515) which
can be used to cross-check the calibration of the exposures from the frame 760l.
This cross-checking shows that the strongest exposure \#3  
has a systematic positive shift from 0.3 \kms\ (positions of \ion{Ca}{ii} $\lambda\lambda3934, 3969$ \AA, 
\ion{Ca}{i} $\lambda4227$ \AA\ at \zabs\ = 0.45207) 
to 0.2 \kms\ (positions of \ion{Fe}{ii} $\lambda\lambda2586, 2600$ \AA\ from the present system) relative to 
the center of each absorption-line system.
In turn, the \ion{Mg}{ii} profiles from the exposure \#8  (700u) 
and \#3 (760l) coincide, i.e., at the position of the \ion{Mg}{ii} lines
these two exposures are coherent. Thus, we can assume that the reference exposure \#8 (700u) has a
positive shift of $\leq 0.2$ \kms.
This gives us a conservative lower limit on the offset of the \ion{Mg}{ii}
lines due to enhanced content of heavy isotopes of 
$\Delta V_{\scriptscriptstyle \rm Mg\,{\sc II}} \ga -0.08$ \kms. 
Assuming 
\mgc/\mgb\ $> 1$, which is supported both by measurements in presolar spinel grains and in
some Milky Way giants (Gyngard \etal\ 2010; Yong \etal\ 2006) and by
theoretical predictions (Karakas \etal\ 2006), we obtain  
(\mgb + \mgc)/\mga $\la 0.7$ (see Fig.~\ref{fg7}, lower panel).
We note that in the previous (lower resolution) spectrum of \object{HE0001--2340} 
the \ion{Mg}{ii} lines from the \zabs\ = 1.5864 system are shifted with respect to
the \ion{Mg}{ii} lines from the present spectrum by $\approx -0.5$ \kms.

\begin{figure*}[t]
\vspace{0.0cm}
\hspace{0.0cm}\psfig{figure=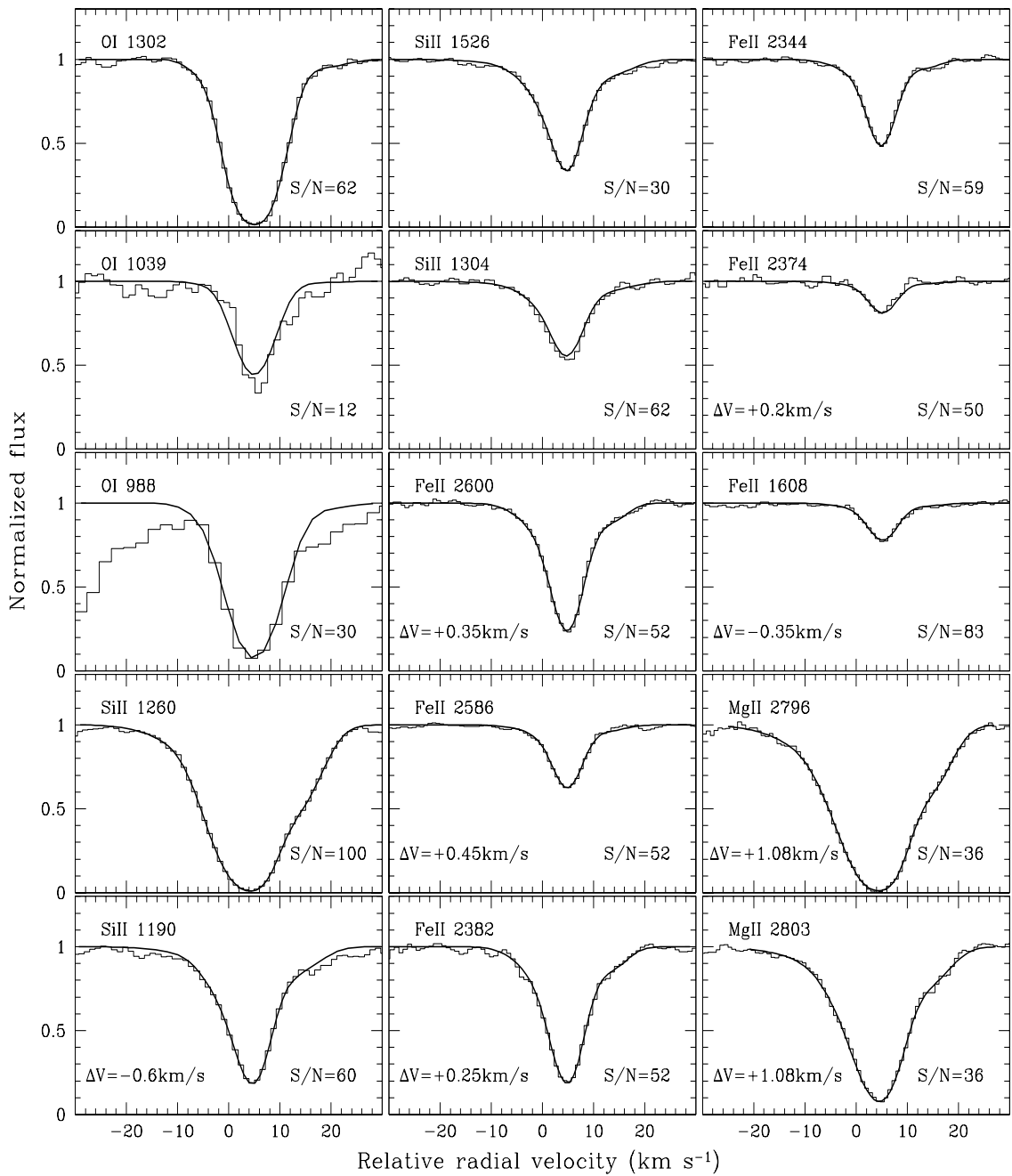,height=18cm,width=16cm}
\vspace{0.5cm}
\caption[]{
Metal absorption lines from
the \zabs = 2.1871 system towards the quasar \object{HE0001--2340}
(solid-line histograms).
Synthetic profiles are plotted by the smooth curves.
The signal-to-noise ratio, S/N, per pixel in the co-added spectrum is indicated in each panel.
Also shown are the relative velocity offsets $\Delta V$: for the
\ion{Si}{ii} $\lambda1190$ \AA\ line and the \ion{Mg}{ii} doublet
they are with respect to the \ion{Si}{ii} $\lambda\lambda1260, 1526$ \AA\ lines, and
for the \ion{Fe}{ii} lines~-- relative to the \ion{Fe}{ii} $\lambda2344$ \AA\ line.
The zero radial velocity is fixed at $z = 2.1871$.
}
\label{fg13}
\end{figure*}

Again, the presence of spectral lines with different
sensitivities to changes in $\alpha$ makes it possible to estimate the value of
$\Delta \alpha/\alpha$, but now with a considerably higher accuracy  
since the errors of the line centers are small.
Namely, by comparing the \ion{Si}{ii} $\lambda1526$ \AA\ line as a transition unaffected by $\alpha$-variations 
with a sensitivity coefficient $Q \approx 0.001$
(Dzuba \etal\ 2002) with the \ion{Fe}{ii} $\lambda\lambda2600, 2586, 2382, 2374, 2344$ \AA\ transitions, 
all having similar sensitivity coefficient of $Q \approx 0.04$ (Porsev \etal\ 2007), we obtain
$\Delta \alpha/\alpha = (-1.5 \pm 2.6)\times 10^{-6}$. 
This is comparable with the most 
stringent constraint on $|\Delta\alpha/\alpha| < 2$ ppm
found from the \ion{Fe}{ii} lines at $z = 1.15$ towards the bright quasar
HE~0515--4414 (Quast \etal\ 2004; Levshakov \etal\ 2005, 2006; Molaro \etal\ 2008b). 
The line \ion{Fe}{ii} $\lambda1608$ \AA\ 
with $Q = -0.016$ is very weak and, besides, its red wing is slightly 
corrupted by the noise spike (Fig.~\ref{fg10}). 
If only the blue (incorrupted) portion of the \ion{Fe}{ii} $\lambda1608$ \AA\ 
profile is taken, then $\Delta \alpha/\alpha$
estimated from the \ion{Fe}{ii} transitions is 
$\Delta \alpha/\alpha = (-0.5 \pm 11.0)\times 10^{-6}$.

\subsection{Absorber at \zabs\ = 1.6515}
\label{sect-4-3}

This system was described in Paper~I where the following relative abundances were
determined: [C/H] = $0.4 - 0.45$, [Mg/C] = $0.1-0.15$, [Fe/C] $\la -0.3$, and [Si/C] = $-0.8$.
Such an abundance pattern relates the absorbing gas with the outflows from AGB-stars. 
A high content of magnesium along with a relative deficit of iron is one of the characteristics
of hot-bottom burning (Herwig 2005) which can result in an enhanced content of the neutron-rich Mg
isotopes. Indeed, analyzing the previous (lower resolution) spectrum of \object{HE0001--2340}
we noticed that \ion{Mg}{ii} lines were offset by $\approx -0.6$ \kms\ relative to the lines 
of other low-ionization species from
this system. Here we present accurate line centers based on the higher resolution spectrum.

The available low-ionization transitions are \ion{C}{ii} $\lambda1334$ \AA, 
\ion{Si}{ii} $\lambda\lambda1260$, 1526 \AA, 
\ion{Fe}{ii} $\lambda2382$ \AA, \ion{Mg}{ii} $\lambda\lambda2796, 2803$ \AA,
and \ion{Mg}{i} $\lambda2852$ \AA. 
However, the \ion{C}{ii} $\lambda1334$ \AA\ and \ion{Si}{ii} $\lambda1260$ \AA\ lines are blended with 
strong Ly-$\alpha$ absorptions and cannot be deconvolved with a sufficiently high accuracy. 
The combined \ion{Mg}{ii} $\lambda\lambda2796, 2803$ \AA\ profiles
consist of 6 co-added exposures (3 from 760l, and 3 from  700u)
aligned with the exposure \#8 (700u).
The weak \ion{Mg}{i} $\lambda2852$ \AA\ line is present only in 3 exposures of
the frame 700u which were simply co-added, i.e., the  profile of this line is mostly
affected by the exposure \#8 with the strongest signal. 
The profiles of \ion{Si}{ii} $\lambda1526$ \AA\ and \ion{Fe}{ii} $\lambda2382$ \AA\ were
prepared as described in Sect.~\ref{sect-3}.

The line centers were determined for the \ion{Si}{ii} $\lambda1526$ \AA, 
\ion{Fe}{ii} $\lambda2382$ \AA, \ion{Mg}{ii} $\lambda\lambda2796, 2803$ \AA, 
and \ion{Mg}{i} $\lambda2852$ \AA\ lines.
Lines of each ion were fitted independently.
The blue wing of \ion{Mg}{ii} $\lambda2796$ \AA\ is blended with
a weak telluric absorption, but the central part of this line is unaffected. 
The fitting parameters are listed in Table~\ref{tbl-6}, the 
observed and synthetic profiles are shown in Fig.~\ref{fg11}. 
The system center calculated as the weighted mean between the centers of the \ion{Si}{ii} $\lambda1526$ \AA\ and 
\ion{Fe}{ii} $\lambda2382$ \AA\ lines is 
$V_c = -4.05\pm0.20$ \kms\ which gives the offsets 
$\Delta V_{\scriptscriptstyle \rm Mg\,{\sc II}} = -0.08 \pm 0.20$ \kms, and
$\Delta V_{\scriptscriptstyle \rm Mg\,{\sc I}} = -0.28 \pm 0.44$ \kms. 

The \ion{Mg}{ii} lines are observed in the range $\Delta \lambda = 7414 - 7434$ \AA.
At both sides of this region there are absorption lines with verified calibration. Namely,
at 7380 \AA, the \ion{Mg}{i} $\lambda2852$ \AA\ line from  the \zabs\ = 1.5864 system is present and
its position coincides with other lines from this system (Table~\ref{tbl-4}). 
At 7472 \AA, we have the \ion{Fe}{ii} $\lambda2344$ \AA\ line from the \zabs\ = 2.1871 
system, and this line is also unshifted (see Sect.~\ref{sect-4-4}). 
Thus, the calibration of the \ion{Mg}{ii} lines is correct.   
As for the \ion{Mg}{i} $\lambda2852$ \AA\ line, it lies close to the \ion{Fe}{ii} $\lambda2374$ \AA\ line 
from the \zabs\ = 2.1871 system which is shifted relative to \ion{Fe}{ii} $\lambda2344$ \AA\
by 0.2 \kms\ (Sect.~\ref{sect-4-4}). 
Therefore, the \ion{Mg}{i} $\lambda2852$ \AA\ line is very probably shifted as well,
i.e., $V_{\scriptscriptstyle \rm Mg\,{\sc I}}$ 
should be\ $-4.13 \pm 0.39$ \kms, and
$\Delta V_{\scriptscriptstyle \rm Mg\,{\sc I}} = -0.08 \pm 0.44$ \kms.  
Because of weakness of 
the \ion{Si}{ii} $\lambda1526$ \AA\ and \ion{Fe}{ii} $\lambda2382$ \AA\ lines 
their positions are measured with large errors making
the shift of Mg lines  statistically indistinguishable from zero.
Under assumption that \mgc/\mgb\ $> 1$, 
a tentative upper limit on (\mgb + \mgc)/\mga is $\la 2.6$.

\begin{table}[t!]
\centering
\caption{Fitting parameters of spectral lines identified in the \zabs\ = 2.1871 absorber.
For each ion, the upper row lists parameters of the main component whereas 
parameters of auxiliary components are listed below.
Indicated are $1\sigma$ statistical errors. } 
\label{tbl-7}
\begin{tabular}{l r@{.}l r@{.}l r@{.}l }
\hline
\hline
\noalign{\smallskip}
\multicolumn{1}{c}{Ion} & \multicolumn{2}{c}{$V$, \kms} & \multicolumn{2}{c}{$b$, \kms} 
& \multicolumn{2}{c}{$N$, \cm} \\
\noalign{\smallskip}
\hline
\noalign{\smallskip}
Fe\,{\sc ii}$^\ast$ & 4&$94\pm0.10$ & 2&$56\pm0.05$ & (6&$12\pm0.04$)E12 \\[-1pt]
           &$2$&$54\pm0.15$ & 6&$47\pm0.17$   & (2&$09\pm0.05$)E12 \\[-1pt]
           & 13&$73\pm0.35$  & 2&$88\pm0.29$ & (4&$06\pm0.14$)E11 \\[2pt]
Si\,{\sc ii} & 4&$92\pm0.09$ & 2&$58\pm0.06$ & (9&$92\pm0.35$)E12 \\[-1pt]
           &$-6$&$6\pm1.0$ & 9&$13\pm0.47$   & (5&$94\pm0.17$)E11 \\[-1pt]
 & 3&$15\pm0.15$  & 6&$57\pm0.05$ & (8&$77\pm0.13$)E12 \\[-1pt]
 & 14&$83\pm0.10$  & 4&$55\pm0.07$ & (1&$2\pm0.1$)E12 \\[2pt]
O\,{\sc i} & 4&$97\pm0.05$ & 4&$49\pm0.03$ & (3&$10\pm0.13$)E14 \\[-1pt]
 & $-6$&$7\pm3.1$ & 1&$0\pm1.0$ & (1&$73\pm0.05$)E12 \\[-1pt]
 & $18$&$9\pm0.5$ & 4&$8\pm2.2$ & (2&$85\pm0.65$)E12 \\[2pt]

Mg\,{\sc ii} & 3&$84\pm0.10$ &  3&$75\pm0.06$ & (7&$76\pm0.14$)E12 \\[-1pt]
 & $-7$&$14\pm0.33$ & 10&$88\pm0.41$ & (7&$10\pm0.26$)E11 \\[-1pt]
 &  $1$&$15\pm0.15$ & 6&$97\pm0.05$ & (4&$56\pm0.05$)E12 \\[-1pt]
 &  13&$67\pm0.15$ & 4&$35\pm0.11$ & (8&$22\pm0.10$)E11 \\[2pt]
\noalign{\smallskip}
\hline
\noalign{\smallskip}
\multicolumn{7}{l}{\footnotesize $^\ast$Lines are aligned with respect to Fe\,{\sc ii} $\lambda2344$ \AA\
(see Fig.~\ref{fg13}).}
\end{tabular}
\end{table}

\subsection{Absorber at \zabs\ = 2.1871}
\label{sect-4-4}

This absorber belongs to an extended sub-DLA system described in detail in Richter \etal\ (2005). It
exhibits multiple lines of \ion{O}{i}, \ion{Si}{ii}, 
\ion{Mg}{ii}, and \ion{Fe}{ii}  
which can be used to check the calibration uncertainties.

The \ion{O}{i} $\lambda\lambda1302$, 1039 \AA, \ion{Si}{ii} $\lambda\lambda1526$, 1304, 1260, 1190 \AA,
\ion{Fe}{ii} $\lambda\lambda2600$, 2586, 2382, 2374, 2344, 1608 \AA, and
\ion{Mg}{ii} $\lambda\lambda2803$, 2796 \AA\ lines are present in many
exposures of different settings. Their profiles were prepared as described in Sect.~\ref{sect-3}.

The \ion{Fe}{ii} $\lambda\lambda2586, 2600$ \AA\ lines are detected in 3 exposures of the
frame 760u and in 3 exposures of the frame 700u (all exposures with different S/N). 
Before co-adding, the exposures were aligned with
the exposure \#3 (760u) showing the highest S/N. 
We note that \ion{Fe}{ii} profiles from this exposure completely
coincide with \ion{Fe}{ii} profiles from the exposure \#8 (700u) which is the highest S/N
exposure in the setting 700u.

The \ion{Fe}{ii} $\lambda2344$ \AA\ line
is present in 3 exposures of 760l and in 3 exposures of  700u.
Again, all of them have different S/N. 
The alignment was performed relative to the exposure \#8 (700u).

The \ion{Fe}{ii} $\lambda\lambda2382, 2374$ \AA\ lines are present only in 3 exposures of 
700u where their profiles are completely consistent, and
the final profiles were obtained by simple co-adding.

The \ion{Mg}{ii} $\lambda\lambda2796, 2803$ \AA\ lines 
are present in 3 exposures of 760u. Before co-adding, 
individual profiles were aligned with the profile from the exposure \#3.

Line profiles of different ions in the \zabs\ = 2.1871 system differ from each other which is clearly seen
in Fig.~\ref{fg12} where several profiles are overplotted. This means that the absorbing gas has
density and velocity gradients. However, all ions trace some central condensation, and we
can estimate the shifts between the lines of different ions using their central components.
Lines of each ion were fitted independently to Gaussians the number of
which was determined as a minimum required to fit all transitions of a given ion with
$\chi^2 \la 1$. The synthetic profiles are shown by solid lines in Fig.~\ref{fg13}.
The fitting parameters are listed in Table~\ref{tbl-7}. 
Some details on the fitting procedure are explained below.

The ion \ion{O}{i} is represented by the lines 1302 \AA\ and 1039 \AA. 
The \ion{O}{i} $\lambda1302$ \AA\ line is almost saturated and the 1039 \AA\ line falls
in a very noisy part of the spectrum. 
To constrain the fitting parameters  
we included in the analysis the \ion{O}{i} $\lambda988$ \AA\ 
line from the previously obtained spectrum of \object{HE0001--2340} with FWHM = 6.8 \kms\
(its profile is also shown in Fig.~\ref{fg13}).

Trial fittings of \ion{Si}{ii} lines showed that the \ion{Si}{ii} $\lambda1190$ \AA\ line 
was noticeably shifted with respect to 
the \ion{Si}{ii} $\lambda\lambda1260, 1526, 1304$ \AA\ lines and
that the synthetic \ion{Si}{ii} $\lambda1304$ \AA\ line came out systematically weaker than the observed one
(see Fig.~\ref{fg13}).
Both the 1190 \AA\ and 1304 \AA\ lines were excluded from the fitting procedure, and 
the fitting parameters were estimated on base of the 1260 \AA\ and 1526 \AA\ lines. 
After the fitting procedure, the synthetic profile of the 1190 \AA\ line was
calculated and the shift of the observed 1190 \AA\ line was evaluated by cross-correlation of 
the synthetic and observed profiles. We found that  \ion{Si}{ii} $\lambda1190$ \AA\ line 
was shifted by $-0.6$ \kms\ relative to the other \ion{Si}{ii} transitions. 
This line is located just at the order edge what probably explains the revealed discrepancy. 
As for the \ion{Si}{ii} $\lambda1304$ \AA\ line, the reason for its intensity to be underestimated whereas the
profiles of the \ion{Si}{ii} $\lambda 1260$, 1190 (shifted), and 1526 \AA\ lines are fitted perfectly, is unclear.
May be it is due to a blend with some unidentified absorption.

Trial fittings of the available \ion{Fe}{ii} lines ($\lambda\lambda2600, 2586, 2374, 2382, 2344, 1608$ \AA) 
revealed that all of them are shifted with respect to each other. 
We are especially interested in calibration of the \ion{Fe}{ii} $\lambda2344$ \AA\ line 
since it lies in the vicinity of the \ion{Mg}{ii} doublet
from the \zabs\ = 1.6515 system (Sect.~\ref{sect-4-3}) and can be used to verify its position.  
We took \ion{Fe}{ii} $\lambda2344$ \AA\ as
a reference line, calculated offsets for the other \ion{Fe}{ii} lines, and aligned all profiles.
Then the aligned \ion{Fe}{ii} lines were fitted together. 
The fitting parameters and the
shifts of the individual lines relative to \ion{Fe}{ii} $\lambda2344$ \AA\ are shown in Table~\ref{tbl-7}. 
The center of the 2344 \AA\ line coincides within the uncertainty interval with the centers of the
\ion{O}{i} and \ion{Si}{ii} lines. Thus, we can conclude that the exposure \#8 from the frame 700u 
(reference exposure for the alignment of the individual \ion{Fe}{ii} $\lambda2344$ \AA\ profiles before co-adding) 
at 7472 \AA\ (position of \ion{Fe}{ii} $\lambda2344$ \AA ) has no offsets
relative to the zero-point. 
Note that there is a good ThAr reference line at 7472 \AA.
However, negative offsets begin to appear at larger wavelengths being
$-0.2$ \kms\ at 7564 \AA\ (\ion{Fe}{ii} $\lambda2374$ \AA ), $-0.25$ \kms\ at 7594 \AA\ 
(\ion{Fe}{ii} $\lambda2382$ \AA ), $-0.45$ \kms\ at 8244 \AA\ (\ion{Fe}{ii} $\lambda2586$ \AA ),
and $-0.35$ \kms\ at 8286 \AA\ (\ion{Fe}{ii} $\lambda2600$ \AA ).

A negative offset is clearly displayed also by the \ion{Mg}{ii} $\lambda\lambda2796, 2803$ \AA\ doublet
located in the range 8912--8936 \AA: it is shifted by $-1.08 \pm 0.13$ \kms\ relative to \ion{Si}{ii} lines
(Table~\ref{tbl-7}).
To what extent this offset is due to a putative high content of heavy Mg isotopes is not clear.  
According to Richter \etal\ (2005),
in the present system [O] $= -1.8$ and [N/O]\ $< -1.5$, i.e., enrichment occurred due to metal-poor SN II.
Thus, from general considerations the enhanced ratio (\mgb + \mgc )/\mga\, 
should not be expected. It is likely that
the shift of the \ion{Mg}{ii} lines 
is mostly caused by the distortions in the calibration process.

This tendency~-- increasing negative offsets with increasing wavelength in the range $\lambda > 7500$ \AA~-- is
detected not only in the present spectrum, but in other UVES spectra as well. 
Such offsets are found both in the archive QSO spectra and in spectra newly obtained within the framework of
the ESO Large Program 185.A-0745.
At the hardware level the UVES is realized by the two arms~-- blue and red, and in the red one there are
two CCD chips~-- lower (REDL) and upper (REDU). 
The blue arm
covers the range 3000~-- 5000 \AA, and the red arm 4200~--11000 \AA. 
The performance of the blue and lower red chips was checked by an
original procedure involving solar radiation reflected by asteroids (Molaro \etal\ 2008a) and
no large systematic shifts in the wavelength calibration between two chips were detected. 
Our data support this result, although they also demonstrate that local line position shifts 
with amplitudes in most cases below $|\Delta V| = 0.4$ \kms\ ($< 1/3$ pixel size) are possible. 
Why the wavelength calibration slides down in spectra
detected with the REDU chip (7500 \AA\ $< \lambda <$ 9000 \AA)
is unclear and this problem needs its thorough investigation. 
In any case it is quite obvious that the studies involving
differential measurements of the line positions may use REDU spectra
only if there are some independent methods to check their wavelength calibration.

This is just the case of the system considered here where calibration 
of \ion{Fe}{ii} $\lambda2344$ \AA\ is verified by comparison with multiple
lines of other ions (\ion{O}{i} and \ion{Si}{ii}). 
Since the sensitivity coefficients of \ion{O}{i} is 
$Q_{\scriptscriptstyle \rm O\,{\sc I}} \approx 0$ (Berengut \& Flambaum 2010)
and of \ion{Si}{ii} is 
$Q_{\scriptscriptstyle \rm Si\,{\sc II}} \approx 0.001$ (Dzuba \etal\ 2002),  
positions of the \ion{O}{i} and \ion{Si}{ii} lines should not be affected 
by putative variations of the fine-structure constant $\alpha$, 
whereas the sensitivity coefficient of the \ion{Fe}{ii} $\lambda2344$ \AA\ transition is 0.036 (Porsev \etal\ 2007).
The difference $\Delta V$ between the \ion{Fe}{ii} $\lambda2344$ \AA\ and \ion{O}{i}/\ion{Si}{ii} line centers
is much smaller than the error of $\Delta V$ 
which is 0.11 \kms\ if the \ion{O}{i} line center is taken as a zero-point,
and 0.13 \kms\ if \ion{Si}{ii} is a reference line. The former
case gives us the value $\Delta \alpha/\alpha = (1.4 \pm 5.1)\times 10^{-6}$, the latter --
$\Delta \alpha/\alpha = (-0.9 \pm 6.0)\times 10^{-6}$. 
In turn, the line \ion{Fe}{ii} $\lambda1608$ \AA\ with $Q = -0.016$ is shifted by 0.3 \kms\ relative to 
the \ion{Fe}{ii} $\lambda2344$ \AA, which
would give us the value $\Delta \alpha/\alpha = (9.6 \pm 4.5)\times 10^{-6}$. This example
confirms the conclusion of Griest \etal\ (2010) that calibration errors 
are the main source of uncertainties in $\Delta \alpha/\alpha$
estimations from absorption lines in quasar spectra.

We note that the present system was used by Chand \etal\ (2004) and
by Murphy \etal\ (2008) to estimate $\Delta \alpha/\alpha$ 
from the transitions \ion{Mg}{ii} $\lambda2803$ \AA, \ion{Al}{ii} $\lambda1670$ \AA, 
\ion{Si}{ii} $\lambda1526$ \AA, and \ion{Fe}{ii} $\lambda\lambda2374, 2586$ \AA\
which were fitted simultaneously using the same model.
It is unclear how this could be done 
taking into account that in the previous \object{HE0001--2340} spectrum 
all peculiarities described above (e.g., extreme shift of the \ion{Mg}{ii} lines and different line profiles of
different ions) were also present.

\section{Conclusions}
\label{sect-5}

In the present paper we attempt to measure the Mg isotope abundances in several absorption systems detected 
in the spectrum of the quasar \object{HE0001--2340}.
The spectrum was obtained with the VLT/UVES with the slit width of 0.7 
arcsec and was read pixel by pixel without binning, thus
ensuring about two times smaller pixel size as was achieved in previous (archived) spectra of this source 
(1.3 \kms\ vs. 2.3 \kms).
The line profiles were prepared individually according to a
special procedure described in Sect.~\ref{sect-3}. Absorption-line systems selected for the study 
reveal multiple lines of the same ions
(e.g., \ion{Si}{ii} $\lambda\lambda1526, 1304, 1260$ \AA, 
\ion{Fe}{ii} $\lambda\lambda2382, 2374, 2344$ \AA\ etc.) 
which can be used to verify the local wavelength calibration. 
The main results are the following.
 
\begin{enumerate}
\item[1.] 
For the first time we measured the abundance of Mg isotopes in the high-redshift
absorber at \zabs\ = 0.45207 which is probably a remnant of the SN Ia explosion of high-metallicity
white dwarf(s). Lines of the \ion{Mg}{ii} doublet are shifted relative to low-ionization transitions
of other ions (\ion{Fe}{i}, \ion{Fe}{ii}, \ion{Ca}{i}, \ion{Ca}{ii}) 
by $\Delta V_{\scriptscriptstyle \rm Mg\,{\sc II}} = -0.44\pm0.05$ \kms, 
and the \ion{Mg}{i} line by $\Delta V_{\scriptscriptstyle \rm Mg\,{\sc I}} = -0.17\pm0.17$ \kms.     
This translates into the isotope abundance ratio \mga:\mgb:\mgc\ 
$= (19\pm11):(22\pm13):(59\pm6)$ 
with strong relative overabundance of heavy Mg
isotopes \mgb+\mgc\ relative to \mga\ as compared to the solar ratio 
\mga:\mgb:\mgc\ = 79:10:11.
\item[2.]
In the absorbers at \zabs\ = 1.5864 and \zabs\ = 1.6515 arising in the gas enriched likely by outflows
from AGB type stars we obtain for the shift of the \ion{Mg}{ii} lines
relative to other ions the values of  
$\Delta V_{\scriptscriptstyle \rm Mg\,{\sc II}} \ga -0.08$ \kms\ at \zabs\ = 1.5864, and 
$\Delta V_{\scriptscriptstyle \rm Mg\,{\sc II}} = -0.08 \pm 0.20$ \kms\ 
at \zabs\ = 1.6515. This allows us to set only upper 
limits on the content of heavy Mg isotopes in the absorbing gas:           
(\mgb + \mgc)/\mga\ $\la 0.7$ 
at \zabs\ = 1.5864, and        
(\mgb + \mgc)/\mga\  $\la 2.6$ at \zabs\ = 1.6515.
\item[3.]
In the present spectrum of \object{HE0001--2340}
the errors of the wavelength calibration 
produce velocity shifts of absorption lines with
amplitudes in most cases below 0.4 \kms\ ($1/3$ pixel size). 
This statement is valid for the wavelength ranges covered by the blue and REDL 
chips of the UVES. However, absorption lines taken with the REDU chip are systematically shifted
toward shorter wavelengths. The discrepancy increases with increasing wavelength and may reach\,
$-1$ \kms\ ($\approx 0.8$ pixel size) in the wavelength interval 
7500 \AA\ $< \lambda <$ 9000 \AA.
\item[4.]
In the absorption system at \zabs\ = 1.5864 with the verified calibration of the metal absorption lines, 
we set a limit on the variation of the fine-structure constant at the level of
$\Delta \alpha/\alpha = (-1.5 \pm 2.6)\times10^{-6}$ 
which is one of the most stringent estimates of this value obtained from optical spectra of QSOs.
\item[5.]
Comparison of statistical errors of the line position measurements
with systematic errors due to miscalibration of the wavelength scale shows
that the systematic error is the main
source of uncertainties in the measurements of
$\Delta \alpha/\alpha$ from quasar absorption-line systems.
\end{enumerate}

\begin{acknowledgements}
The project has been supported in part by
the RFBR grants 09-02-12223 and 09-02-00352, 
by the Federal Agency for Science and Innovations grant
NSh-3769.2010.2,
and by the Chinese Academy of Sciences visiting professorship
for senior international scientists grant No. 2009J2-6. 
HJL is supported by the NSF of China Key Project No. 10833005, the Group Innovation
Project No. 10821302, and by 973 program No. 2007CB815402.
\end{acknowledgements}

\end{document}